\documentclass[a4paper,11pt]{article}
\usepackage{jcappub} %
\usepackage{lineno} 
\usepackage{appendix}
\usepackage{orcidlink}
\usepackage{mathtools}
\usepackage{bm}

\usepackage{enumitem}
\usepackage{soul}
\usepackage[normalem]{ulem}

\newcommand{\beq}{\begin{equation}}
\newcommand{\eeq}{\end{equation}}

\newcommand{\beqn}{\begin{eqnarray}}
\newcommand{\eeqn}{\end{eqnarray}}

\usepackage{tikz}

\usepackage{footnote}

\usepackage[mathscr]{euscript}
\usepackage{comment}

\newcommand{\Mp}{M_{\mathrm{Pl}}}

\begin{document}
\title{
 Phenomenology of
Inflaton-Driven Early QCD
Confinement and Solution to Axion Isocurvature Problem
}

\author[a,b]{Evangelos I. Sfakianakis,}\author[a,c]{ Barmak Shams Es Haghi} \author[a,d,e]{and Katherine Freese}

\affiliation[a]{Texas Center for Cosmology and Astroparticle Physics, Weinberg Institute for Theoretical Physics, Department of Physics, The University of Texas at Austin, Austin, TX, 78712, USA}
\affiliation[b]{Department of Physics, Harvard University, Cambridge, MA, 02131, USA}

\affiliation[c]{Department of Physics and Astronomy, University of Utah, Salt Lake City, UT, 84112, USA}

\affiliation[d]{The Oskar Klein Centre, Department of Physics, Stockholm University, AlbaNova, SE-10691 Stockholm, Sweden}
\affiliation[e]{Nordic Institute for Theoretical Physics (NORDITA), 106 91 Stockholm, Sweden}

\emailAdd{evangelos.sfakianakis@austin.utexas.edu}
\emailAdd{barmak.shams@utah.edu}
\emailAdd{ktfreese@utexas.edu}

\abstract{
    We study the phenomenology of early QCD confinement during inflation, driven
by a direct coupling between the inflaton and Standard Model gluons. This coupling dy-
namically raises the QCD confinement scale, making the axion sufficiently heavy to suppress
isocurvature perturbations during the CMB epoch. As inflation proceeds, the confinement
scale decreases and the axion becomes light, allowing de Sitter fluctuations during the late
stages of inflation or post-inflationary thermal fluctuations to generate the observed dark
matter abundance. In addition, QCD-induced corrections to the inflationary potential can shift the scalar spectral index towards smaller values, providing a further observational handle. We embed this mechanism in an $\alpha$-attractor model of inflation and explore the resulting parameter space. We show that, in the minimal scenario with reheating
into gluons, successful dark matter production requires deconfinement to occur shortly after the CMB window. Extensions involving reheating through heavy right-handed neutrinos
generally require large Yukawa couplings, which induce sizable loop corrections that spoil
inflationary dynamics. We show that this tension can be resolved in the presence of supersymmetry and derive constraints on the SUSY breaking scale that allow the mechanism
to remain viable within plateau models of inflation. Treating the reheat temperature as a
free parameter further enlarges the viable parameter space. 
}

\maketitle

\section{Introduction}

In this work, we study in detail a mechanism for solving the QCD axion isocurvature problem.  This paper serves as a companion to a shorter paper presented recently~\cite{Freese:2026xax}.  We begin with a discussion of QCD axions and the isocurvature problem that accompanies their formation in a pre-inflationary scenario, and then turn to a proposed resolution, which we then study in detail for the example of $\alpha$-attractor inflation models.

The Peccei-Quinn (PQ) solution~\cite{Peccei:1977hh,Peccei:1977ur} to the strong CP problem introduces a global U(1)$_{\rm PQ}$ symmetry that is spontaneously broken at a high energy scale $f_a$. This symmetry breaking gives rise to the axion, the associated pseudo–Nambu–Goldstone boson~\cite{Weinberg:1977ma,Wilczek:1977pj}, which is an excellent dark matter (DM) candidate~\cite{Zhitnitsky:1980tq, Dine:1981rt,Preskill:1982cy,Abbott:1982af,Dine:1982ah}. 
Since the PQ symmetry is anomalous with respect to SU(3)$_{\rm C}$, nonperturbative QCD effects explicitly break it and generate (approximately) a cosine potential, thereby giving the axion a mass. The axion potential has a characteristic scale of order $\Lambda^4_{\rm QCD}$ which leads to an axion mass of order \beq
\label{eq:axionmassdef}
m_a\sim { \Lambda^2_{\rm QCD}\over f_a} \, ,\eeq where $\Lambda_{\rm QCD}$ is the QCD confinement scale. The relic abundance of the axion is determined by the misalignment mechanism~\cite{Preskill:1982cy,Abbott:1982af,Dine:1982ah}, in which the axion field undergoes coherent oscillations about the minimum of its potential after starting from an initial displacement.
The resulting axion cosmology depends on whether the PQ symmetry is broken before or after a period of inlation in the early universe (for a recent review, see Ref.~\cite{OHare:2024nmr}).  
In the pre-inflationary scenario, a single initial displacement of the axion field is selected and stretched to super-horizon scales, producing a homogeneous initial misalignment angle across the observable Universe. Since the relic abundance of the axion set by the misalignment mechanism mainly depends on the initial field displacement and the axion mass, the freedom in choosing the initial displacement allows for a wide possible range of axion masses consistent with observed relic abundance of DM.  This freedom, however, comes at the price of accepting a certain degree of fine-tuning in the initial conditions. In the post-inflationary case, in which the PQ symmetry either breaks after inflation or is restored by reheating, the axion misalignment angle takes random values with equal probability across different patches of the Universe and is therefore effectively set by an average dimensionless value of ${\cal O}(1)$, seemingly making the axion mass predictive. However, this predictivity is spoiled by axion emission from topological defects; after the PQ symmetry is ultimately broken, the axion abundance also receives contributions from topological defects such as cosmic strings and domain walls~\cite{Hogan:1988mp,Kolb:1993zz}, whose formation and decay must be properly taken into account (for a recent study, see Ref.~\cite{Benabou:2024msj}). 

In this paper we focus on the pre-inflationary scenario, which 
 requires PQ symmetry breaking to take place at an energy scale above the Hubble scale during inflation, as well as above the reheating temperature. This case typically suffers from overproduction of isocurvature axion fluctuations --- a problem that the model of this paper aims  to resolve.
In the case where the axion field exists during inflation, it obtains quantum fluctuations which
 are stretched to cosmological scales. Later, when the axion acquires its mass, they turn into cold DM isocurvature perturbations~\cite{Axenides:1983hj,Lyth:1989pb,Turner:1990uz,Linde:1991km}, leaving imprints on the cosmic microwave background (CMB) temperature anisotropies.
 These perturbations are, however, observationally constrained to be very small. The stringent constraint on the amplitude of the isocurvature power spectrum measured by {\it Planck}~\cite{Planck:2018jri} imposes an upper bound on the Hubble scale during inflation which is in tension with a high-scale inflationary models. 
 
In this work, we study in detail a mechanism for solving the axion isocurvature problem, proposed independently in  Refs.~\cite{Freese:2026xax, Dvali:2026ceb}.   
The mechanism is based on  a coupling between the inflaton and gluons, which can dynamically raise the QCD confinement scale during inflation, making the axion sufficiently heavy to suppress isocurvature perturbations during the part of inflation where modes are produced that are detectable in the CMB, typically 50-60 $e$-folds before the end of inflation. As inflation proceeds, the QCD scale decreases, leading to a deconfinement transition that renders the axion light again. The resulting late-time fluctuations can account for the observed dark matter abundance.

A simple explanation of the axion isocurvature problem is as follows (with a more detailed treatment provided in Section~\ref{sec:deSitter}). In the standard pre-inflationary scenario, the axion is a  light spectator field during inflation. Much like the inflaton field that generates the density fluctuations responsible for large-scale structure~\cite{Starobinsky:1994bd}, the axion receives quantum fluctuations of order $\delta a \sim H_I / 2\pi$, where $H_I$ is the inflationary Hubble scale. As the universe expands, these fluctuations are stretched to super-horizon scales, creating small, spatially varying differences in the axion field value across different Hubble patches. Later in the universe's history, when the axion acquires its mass and begins to oscillate, these spatial field differences translate directly into spatial variations in the axion dark matter density. Because these density fluctuations are generated independently of the radiation bath, they are known as isocurvature perturbations. If the energy scale of inflation is high, the quantum kicks $\delta a$ are large, producing isocurvature perturbations that significantly exceed the tight upper bounds set by CMB observations.

A variety of ways have been proposed to suppress axion isocurvature perturbations: the radial component of the PQ symmetry breaking scalar field attaining values larger than $f_a$~\cite{Linde:1990yj,Linde:1991km}, a non-minimal
derivative coupling of the QCD axion to gravity~\cite{Folkerts:2013tua,Berbig:2024ufe}, hidden sector magnetic monopoles~\cite{Nomura:2015xil,Kawasaki:2017xwt}, or the axion acquiring a heavy mass during inflation. This heavy mass may arise from a high-scale QCD confinement (much higher than the standard confinement near $\sim 400\,{\rm MeV}$) triggered by a sufficiently large expectation value of the Higgs field during inflation~\cite{Choi:1996fs,Jeong:2013xta,Takahashi:2015waa,Choi:2015zra,Koutsangelas:2022lte}, PQ symmetry breaking by hidden gauge interactions~\cite{Takahashi:2015waa}, or from promoting the strong coupling to a dynamical field-dependent quantity, where the large expectation value of a scalar field during inflation induces early QCD confinement~\cite{Dvali:1995ce}.

Here we focus on a mechanism to escape the problem of ``too large axion isocurvature fluctuations" in which
 the axion acquires a large mass during inflation. In that case, quantum fluctuations cannot drive the field value far from the minimum of the PQ potential at $\langle a_I \rangle = 0$, so that all Hubble patches at the end of inflation have roughly the same value of the axion field and no isocurvature perturbations result.  Indeed, that is the approach taken by a variety of previous authors, as well as in this work. 
The large axion mass during inflation must of course transition to a light axion mass appropriate for axion dark matter production. A way to get a time dependent axion mass is via a dynamical confinement scale $\Lambda_{\rm QCD}$,
since $m_a \propto \Lambda_{\rm QCD}^2 /f_a$ where $f_a$ is the axion decay constant.  Thus a solution to the axion isocurvature problem takes the confinement scale and hence the axion mass to be large during inflation while both become much smaller after inflation. A number of approaches that render the confinement scale dynamical have been proposed, including using a scalar field coupled to gluons (via $\phi G_{\mu\nu} G^{\mu\nu}$).  Only recently has the inflaton been used as the scalar field that gives rise to a dynamical $\Lambda_{\rm QCD}$~\cite{Freese:2026xax, Dvali:2026ceb}. The dynamics analyzed here  will generate a sufficiently large axion mass during inflation that suppresses isocurvature perturbations, while a subsequent transition to a light axion allows for the generation of de Sitter fluctuations and/or thermal fluctuations during reheating, which give rise to dark matter production.

The cosmological implications of an early QCD confinement arising from a dynamical strong coupling have been explored in Ref.~\cite{Ipek:2018lhm}, while its impact on axion DM abundance has been studied in Ref.~\cite{Heurtier:2021rko}. The underlying idea is the presence of a Standard Model (SM) singlet scalar field which has a coupling to the gluons induced by a non-renormalizable operator. This scalar field, when its expected value is non-zero, renormalizes the gluon wave function and consequently modifies the effective strong coupling and its running~\cite{Ipek:2018lhm}.

In this work, we explore in detail the consequences of  the inflaton itself being the scalar  field responsible for making the strong coupling dynamical. Since this scalar is not protected against quantum corrections, the inflaton potential must dominate over the back-reaction induced by its coupling to gluons. As the inflaton evolves under its potential, its field value induces an early QCD confinement at high scales, thereby causing the axion to be heavy during CMB-relevant epochs and consistent with isocurvature bounds. Throughout inflation, the inflaton rolls toward the minimum of its potential, causing the QCD confinement scale to decrease until it eventually falls below the Hubble scale before the end of inflation. At that point QCD deconfines, and the axion becomes light and acquires either de Sitter fluctuations generated after the CMB-relevant modes have exited the horizon or thermal fluctuations produced during reheating. These fluctuations ultimately lead to an axion population that behaves as cold DM.

To make this mechanism concrete, we employ an $\alpha$-attractor potential that yields an almost constant Hubble scale during inflation. 
In Ref.~\cite{Freese:2026xax} we have shown that this mechanism is compatible with plateau inflationary potentials, including $\alpha$-attractor-like models.
Within this framework,  requiring a successful inflationary period consistent with observations, together with axion isocurvature constraints and theoretical consistency, imposes a hierarchy among the relevant energy scales, which in turn restricts the viable parameter space. The built-in interaction between the inflaton and gluons provides a natural reheating channel in this model. As we show, if this is the only reheating channel, the model remains viable only within a narrow region of the parameter space. However, by introducing an additional reheating channel through a Yukawa coupling between the inflaton and heavy right-handed neutrinos, the viable parameter space significantly expands.
Finally, if the reheat temperature is treated as a free parameter rather than tied to a specific reheating mechanism, the viable parameter space is enlarged and the QCD sector can induce a modest shift in the inflationary slow-roll parameters. This provides an additional phenomenological handle on the model, since the scalar spectral index $n_s$ is measured with high precision and small shifts in its predicted value can be relevant for comparisons with CMB data.

The outline of this paper is as follows.
In Section~\ref{sec:model}, we present the general framework of our model, introducing the inflaton-gluon coupling, discussing the modification of the QCD confinement scale during inflation, and analyzing axion production from de Sitter fluctuations and reheating, along with the resulting dark matter abundance. 
In Section~\ref{sec:realization}, we embed this mechanism in a concrete inflationary realization based on $\alpha$-attractors and examine the associated constraints. Section~\ref{sec:results} contains our main results:
we first present the parameter-space analysis of the minimal scenario and then explore extensions, including different reheating channels,  as well as the possibility for detectable shifts in $n_s$.  In Section~\ref{sec:GW}, we briefly discuss the gravitational-wave signal associated with the confinement–deconfinement transition during inflation. We conclude in Section~\ref{sec:conclusion} with a summary of our findings and directions for future work. 

Before proceeding, we summarize some of the key notation used throughout this work.
We denote by $N$ the number of $e$-folds before the end of inflation, with $N=0$ at the end of inflation and $N \simeq 50$--$60$ corresponding to CMB scales.
We define $N_{\rm dec}$ as the number of $e$-folds before the end of inflation at which the QCD confinement scale drops below the Hubble scale, triggering deconfinement.
The quantity $N_{\rm dec}$ is also the number of $e$-folds during which the axion is light and accumulates de Sitter fluctuations.
The axion field is written as $a = f_a \theta$, where $\theta$ is the dimensionless misalignment angle, and we denote by $\sigma_\theta^2$ the variance of $\theta$ arising from de Sitter and thermal fluctuations.
We use the reduced Planck mass $M_{\rm {Pl}} \simeq 2.4 \times 10^{18}$ GeV throughout this work.

\section{Model}
\label{sec:model}

To avoid the axion isocurvature problem, we desire a large axion mass during inflation, to avoid production of large amplitude space-dependent quantum fluctuations; subsequently the mass drops to its current small value, allowing for dark matter production.
The way our model accomplishes this transition is via a dynamical confinement scale, which is controlled by the coupling of gluons to the inflaton field.

We begin by describing our inflationary setup and examine how the coupling between the inflaton field and gluons can make QCD confinement dynamical, along with the resulting implications for the QCD axion mass, extending the discussion of Ref.~\cite{Freese:2026xax}. We next discuss axion isocurvature perturbations, emphasizing how a mass exceeding the Hubble scale during inflation can suppress them. Next, we discuss the possible reheating channels in our framework and highlight the role of the reheating temperature in our analysis. We subsequently analyze the axion DM abundance arising from both de Sitter (quantum) and thermal fluctuations. Finally, we outline the conditions required for suppressing QCD axion isocurvature modes in our setup, which in turn implies a hierarchy among the relevant scales in the scenario.

\subsection{Dynamical QCD scale driven by coupling to the Inflaton Field}

We consider a simple single-field inflationary model with potential $V(\phi)$. The details of the model are not expected to  significantly change our results, as long as there exists a plateau~\cite{Freese:2026xax}, where the Hubble scale remains relatively constant during inflation.\footnote{This mechanism with the coupling of Eq.~\eqref{eq:inflatongluoncoupling} does not work for generic monomial  inflationary potentials~\cite{Freese:2026xax}. However, CMB data point towards plateau models through the small preferred values of the tensor to scalar ratio, making our construction well motivated.} The inflaton Lagrangian, using the mostly plus metric signature, is 
\beq
{\cal L}_{\rm {infl}} =
-{1\over 2}\partial_\mu\phi\partial^\mu\phi - V(\phi) \, .
\label{eq:interactionterm}
\eeq
We introduce an  interaction term between the inflaton and the SM gluons, as described in Ref.~\cite{Ipek:2018lhm} for a generic scalar field.\footnote{Ref.~\cite{Dvali:2026ceb} appeared shortly before our companion paper~\cite{Freese:2026xax} describing a similar inflationary mechanism.} The inflaton-gluon interaction changes the Lagrangian for the gluon field strength from ${\cal L}_g^{\rm{SM}} = -G_{\mu\nu}G^{\mu\nu}/(4g_{s0}^2)$ to
\begin{equation}
  {\cal L}_g  = -{1\over 4}\left ({1\over g_{s0}^2} +{ \phi\over M} \right ) G_{\mu\nu}G^{\mu\nu}
 \equiv {\cal L}_g^{\rm{SM}} + {\cal L}_{\rm{int}} \, .\label{eq:inflatongluoncoupling}
\end{equation}
We define $g_{s0}$ to be the QCD coupling for vanishing value of the $\phi$ field. The mass parameter $M$ denotes the scale of a non-renormalizable interaction
between the inflaton $\phi$ and the SM gluons, arising, for example, from integrating out heavy colored fermions.
The full action of our model thus becomes
\beq{\cal S} = \int d^4x \sqrt{-g} \left (
{\cal L}_{\rm {infl}} + {\cal L}_{\rm{SM}}+{\cal L}_{\rm {int}}
\right ) \, .
\eeq

In general, a nonzero average value of the inflaton field,  denoted by $\langle \phi \rangle$, renormalizes the gluon kinetic term and thereby modifies QCD dynamics. 
Adding the effect of $\langle \phi\rangle$ to the usual running of the QCD coupling leads to an equation for the strong coupling $\alpha_s$~\cite{Ipek:2018lhm}:
\beq
{1\over \alpha_s(\mu,\langle\phi\rangle)} = {33-2n_f\over 6\pi} \ln \left (
\mu\over \Lambda_0
\right ) +4\pi{ \langle \phi\rangle\over M} \, ,
\label{eq:alphas}
\eeq
where  $n_f$ is the number of active quarks (those with masses smaller than the scale $\mu$) and $\Lambda_0$ is the standard QCD scale at $\langle\phi\rangle =0$. Confinement occurs when $\alpha_s^{-1}\simeq 0$ and therefore a value $\langle \phi\rangle<0$
leads to QCD confinement occurring at a scale $\mu = \Lambda\left (\langle \phi \rangle \right )$ which can be much higher than $\Lambda_0\simeq 400$~MeV.\footnote{The QCD scale is not a universal constant and depends on the
renormalization scheme, loop order, and number of active flavors. We follow the
conventions of Ref.~\cite{Heurtier:2021rko} and take
$\Lambda_{0}\simeq 400\,\mathrm{MeV}$. Due to the exponential sensitivity of $\Lambda(\bar\phi)$ on $\bar\phi/M$, a difference choice of $\Lambda_0$ can be offset by a minute change of $M$.} During inflation, the average value of the inflaton field corresponds to the homogeneous classical inflaton background, which we denote by $\bar{\phi}$, so that $\langle \phi \rangle = \bar{\phi}$. 
Setting the left-hand side of Eq.~\eqref{eq:alphas} to zero, we can solve for $\mu = \Lambda(\langle \phi \rangle ) = \Lambda(\bar\phi)$.
Confinement occurs at
\begin{equation}
    \Lambda(\bar\phi )
    = \Lambda_0 \exp \left [{{24\pi^2\over 2n_f-33}{\bar\phi\over M}}\right]
    \simeq  \Lambda_0 e^{-11 \bar\phi/M},
    \label{eq:LambdaQCD}
\end{equation}
where we used $n_f=6$ in the last step. 
The QCD scale $\Lambda_{\rm QCD} = \Lambda(\bar\phi)$ in our model depends on the value of the inflaton VEV and goes to the usual value when the inflaton goes to zero.\footnote{In Section~\ref{sec:realization} we discuss  the conditions for $\Lambda(\bar\phi)\to\Lambda_0$  and argue that it occurs immediately after inflation.} 
We will take $\bar \phi < 0$ during the inflationary epoch (with the field rolling towards $\phi=0$), so that the QCD confinement scale during inflation is much larger than its value today.\footnote{To ensure the gluon kinetic term remains positive-definite and avoids a ghost instability, we require $\frac{1}{g_{s0}^2} + \frac{\phi}{M} > 0$. Because $\phi < 0$ and rolls toward the origin during inflation, this prefactor of Eq.~\eqref{eq:inflatongluoncoupling}  increases over time; therefore, it is sufficient that this condition holds at the onset of the observable 60 $e$-folds.}

As the inflaton-dependent coupling shifts the confinement scale, all nonperturbative
QCD observables become functions of $\Lambda\left (\bar\phi\right )$, including the dynamics
responsible for generating the axion potential.  Since in our setup the inflaton VEV modifies the confinement scale to
$\Lambda_{\rm QCD}=\Lambda(\bar\phi)$, the axion mass during inflation is correspondingly
enhanced, as per Eq.~\eqref{eq:axionmassdef}.  In this framework, where $\Lambda_{\rm QCD}$ can be exponentially
larger during inflation than in the present Universe, this effect becomes a central ingredient
in the phenomenological implications discussed below. 
We define $N_{\rm dec}$ as the number of $e$-folds before the end of inflation at which the QCD confinement scale drops below the Hubble scale, triggering deconfinement.

\subsection{Axion production from de Sitter fluctuations and isocurvature modes}
\label{sec:deSitter}

If the PQ symmetry is broken before or during inflation and is not restored afterward, the axion field is homogenized over the observable universe, but acquires quantum fluctuations during inflation. These fluctuations generically source  DM isocurvature perturbations (see e.g.~Ref.~\cite{Allali:2025pja}), which are tightly constrained by CMB observations. In this section we review the generation of axion isocurvature perturbations and show explicitly how an axion mass which is  larger than the Hubble scale during inflation suppresses them. Furthermore, we show how even a limited time before the end of inflation where the axion field is light can generate a significant amount of axion DM particles, without producing observable  isocurvature modes in the CMB, due to the different length-scales involved.

 We parameterize the axion field as
\begin{equation}
a(x) = f_a \, \theta(x),
\end{equation}
where $f_a$ is the axion decay constant and $\theta$ is the misalignment angle.
Once the axion begins to oscillate in an approximately quadratic potential,
its energy density scales as $\rho_a \propto m_a^2 a^2 \propto m_a^2 f_a^2 \theta^2$.
For small fluctuations $\delta \theta$ around a local homogeneous background value $\theta_{\rm bg}$, the relative energy density fluctuations are
\begin{equation}
\frac{\delta \rho_a}{\rho_a} \simeq 2 \frac{\delta \theta}{\theta_{\rm bg}},
\qquad (|\delta\theta|\ll|\theta_{\rm bg}|).
\end{equation}

Since the axion energy density is negligible during inflation, its fluctuations do not
contribute appreciably to the total curvature perturbation at that time.
The axion--radiation isocurvature perturbation is defined as
\begin{equation}
S_{a\gamma} \equiv 3(\zeta_a - \zeta_\gamma)
= \frac{\delta\rho_a}{\rho_a} - \frac{3}{4}\frac{\delta\rho_\gamma}{\rho_\gamma}.
\end{equation}
On superhorizon scales, and on uniform-radiation hypersurfaces ($\delta\rho_\gamma=0$),
this reduces to $S_{a\gamma}\simeq \delta\rho_a/\rho_a \simeq 2\,\delta\theta/\theta_{\rm bg}$.
This means that given a power spectrum ${\cal P}_{\theta}$ for $\theta$, the power spectrum for the isocurvature perturbations is ${\cal P}_S = 4 {\cal P}_{\theta}/\theta_{\rm bg}^2$.

We quantize the axion field as usual in conformal time $d\tau = dt/a$:
\begin{equation}
a(\tau,\mathbf{x}) = \int \frac{d^3 k}{(2\pi)^3} \left[ u_k(\tau)\,\hat a_{\mathbf{k}}\,e^{i\mathbf{k}\cdot\mathbf{x}} + u_k^*(\tau)\,\hat a_{\mathbf{k}}^\dagger\,e^{-i\mathbf{k}\cdot\mathbf{x}} \right],
\end{equation}
with the standard commutation relation $[\hat a_{\mathbf{k}}, \hat a_{\mathbf{k}'}^\dagger] = (2\pi)^3 \delta^{(3)}(\mathbf{k}-\mathbf{k}')$.

The wave-function of the axion field for $m_a\ll H$ in de Sitter space is
\begin{equation}
    u_k(\tau) = (i-k\tau) {H\over 2k^3}e^{-ik\tau} \, .
    \label{eq:ukmassless}
\end{equation}
If we quantize the dimensionless field $\theta(\tau,\mathbf{x})$, the mode-function is multiplied by $f_a^{-1}$.
After horizon crossing ($k\tau =1$), the modes freeze, leading to the corresponding power spectrum:
\begin{equation}
\mathcal{P}_S(k) \simeq \frac{4}{\theta_{\rm bg}^2} \mathcal{P}_\theta(k) = \frac{4}{\theta_{\rm bg}^2}  \left( \frac{H_k}{2\pi f_a} \right)^2 \, ,
\end{equation}
where \(\mathcal{P}_\theta(k)\) is the power spectrum of axion field fluctuations and $H_k$ is the Hubble scale at the time the mode $k$ crosses the horizon. 

In scenarios where the PQ symmetry is broken before inflation, the ensemble average of the misalignment angle vanishes, $\langle \theta \rangle = 0$. In this case, we characterize the typical background value in a given Hubble patch by the root-mean-square (rms) of the stochastic distribution, $\theta_{\rm bg} \to \theta_{\rm rms} \equiv \sqrt{\langle \theta^2 \rangle}$. The axion isocurvature power spectrum is then written as before with the substitution $\theta_{\rm bg}^2\to \langle\theta^2\rangle$.

We evaluate these fluctuations at the CMB pivot scale, $k_* = 0.05 \text{ Mpc}^{-1}$. The isocurvature fraction at this scale is defined as:
\begin{equation}
\beta_{\rm iso}(k_*) \equiv \frac{\mathcal{P}_S(k_*)}{\mathcal{P}_\zeta(k_*) + \mathcal{P}_S(k_*)} \simeq \frac{\mathcal{P}_S(k_*)}{\mathcal{P}_\zeta(k_*)},
\end{equation}
where the last equality holds because isocurvature modes are highly subdominant ($\mathcal{P}_S \ll \mathcal{P}_\zeta$), and the amplitude of the curvature power spectrum is measured to be $\mathcal{P}_\zeta \simeq 2.1 \times 10^{-9}$. Current CMB data constrain uncorrelated isocurvature perturbations to be $\beta_{\rm iso}(k_*) <0.038$~\cite{Planck:2018vyg}, which implies an upper bound on the Hubble scale $H_*$ at the time the CMB pivot scale exits the horizon:
\begin{equation}
H_* \lesssim \pi f_a \theta_{\rm rms} \sqrt{\beta_{\rm iso}^{\rm max} \, \mathcal{P}_\zeta}.
\end{equation}
For \(f_a \sim 10^{12}\,\mathrm{GeV}\) and \(\theta_{\rm rms} \sim \mathcal{O}(1)\), this bound typically requires \(H_* < 10^{8}\,\mathrm{GeV}\), in strong tension with high-scale inflation.

When the axion (scalar field) mass during inflation cannot be neglected, the  positive-frequency mode, which connects to the Bunch-Davies (BD) vacuum at early times, is
\begin{equation}
u_k(\tau)=
\frac{\sqrt{\pi}}{2}\,e^{i(\nu+\frac12)\frac{\pi}{2}}\,
H\,(-\tau)^{3/2}\,
H_{\nu}^{(1)}(-k\tau) \, ,
\qquad
\nu\equiv\sqrt{\frac{9}{4}-\frac{m^2}{H^2}}.
\label{eq:uk_hankel}
\end{equation}
For a massless field \(\nu=3/2\) reproduces the result of Eq.~\eqref{eq:ukmassless}.
When the   mass satisfies $m_A>3H/2$, it is more convenient to write Eq.~\eqref{eq:uk_hankel} as 
\beq
\label{eq:BD_heavy_prefactor}
u_k(\tau)
=
\frac{\sqrt{\pi}}{2}\,e^{i\pi\over 4}\,e^{-\pi\tilde\nu\over 2}\,
H\,(-\tau)^{3/2}\,
H_{i\tilde\nu}^{(1)}(-k\tau)
 \, ,\qquad \tilde\nu= -i\nu = \sqrt{\frac{m^2}{H^2}-\frac{9}{4}}\;>\;0.
\eeq
A useful  asymptotic expansion of the Hankel function, valid for $\tilde \nu\gg x^2$
is
\begin{equation}
H^{(1)}_{i\mu}(x)
\;\xrightarrow[\mu\gg x^2]{}\;
\sqrt{\frac{2}{\pi\mu}}\;
e^{+\pi\mu/2}\;
\left(\frac{e\,x}{2\mu}\right)^{i\mu}
\left[1+\mathcal{O}\!\left(\frac{x^2}{\mu}\right)+\mathcal{O}\!\left(\frac{1}{\mu}\right)\right].
\label{eq:Hankel_large_mu}
\end{equation}
Using
\[
\left(\frac{e\,x}{2\mu}\right)^{i\mu}
=
\left(\frac{x}{2}\right)^{i\mu}\,e^{\,i\mu(1-\log\mu)},
\]
we can rewrite Eq.~\eqref{eq:Hankel_large_mu} as
\begin{equation}
H^{(1)}_{i\mu}(x)
\;\xrightarrow[\mu\gg x^2]{}\;
\sqrt{\frac{2}{\pi\mu}}\;
e^{+\pi\mu/2}\;
\left(\frac{x}{2}\right)^{i\mu}\,
e^{\,i\mu(1-\log\mu)}\,
\Big[1+\cdots\Big].
\label{eq:Hankel_large_mu_rewrite}
\end{equation}
Substituting \eqref{eq:Hankel_large_mu_rewrite} into \eqref{eq:BD_heavy_prefactor} gives
\begin{align}
u_k(\tau,\mu)
&\simeq
H_I\sqrt{\frac{\pi}{2}}\;e^{i\pi/4}\,e^{-\pi\mu/2}\,(-\tau)^{3/2}
\left[
\sqrt{\frac{2}{\pi\mu}}\;e^{+\pi\mu/2}\;
\left(\frac{x}{2}\right)^{i\mu}e^{\,i\mu(1-\log\mu)}
\right]
\nonumber\\[4pt]
&=
H_I\,\sqrt{\frac{1}{\mu}}\;e^{i\pi/4}\,(-\tau)^{3/2}\,
\left(\frac{-k\tau}{2}\right)^{i\mu}\,
e^{\,i\mu(1-\log\mu)}.
\label{eq:BD_heavy_asymp_cancel}
\end{align}
Thus, at late times, the amplitude of the mode-function for heavy fields decays exponentially since $(-\tau)^{2/3}\sim a^{-3/2}\sim e^{-3Ht/2}\sim e^{-3N/2} $. This is  in contrast to the massless (or light) axion case, where a ``frozen" mode exists that can source significant isocurvature fluctuations.

\subsubsection{Evolving axion mass: a solution to the isocurvature problem}

\begin{figure}
    \centering
\includegraphics[width=0.5\linewidth]{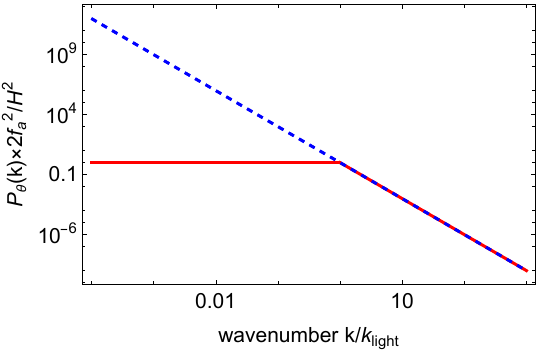}
    \caption{
    The isocurvature power spectrum based on the broken power law of Eq.~\eqref{eq:Pthetasplit} (red), corresponding to a heavy axion at early times and a light axion at late times during inflation.
   The wavenumber corresponding to the time of  the transition between  $m\gtrless 3H/2$ is denoted as $k_{\rm light}$.
    For comparison, the power spectrum of a massless spectator field is given in blue-dashed. The suppression for $k<k_{\rm light}$ is evident.
     }
    \label{fig:uk}
\end{figure}

For a mass that changes  during inflation, interpolating between $m>3H/2$ at early times and $m<3H/2$ at late times the power spectrum can be approximated by a broken power law (see Ref.~\cite{Chakraborty:2025lyp})
\begin{equation}
\begin{split}
P_\theta(k) &= {H^2 \over 2f_a^2} {1\over k_*^3}\, ,\quad k\ll k_{\rm light} \\
P_\theta(k) &= {H^2 \over 2f_a^2} {1\over k^3}\, ,\quad k\gg k_{\rm light} \, ,
\end{split}
\label{eq:Pthetasplit}
\end{equation}
 where the wavenumber $k_{\rm light}$ encodes  the time of the mass transition between $m\gtrless 3H/2$. Figure~\ref{fig:uk} shows the suppression of the axion power spectrum due to a large mass.

 In our framework the wavenumber $k_{\rm light}$, encoding the time of the mass transition, corresponds precisely to the time of QCD deconfinement, $N_{\rm dec}$. The physical requirement for early QCD confinement—and thus for the axion to be heavy ($m_a > 3H/2$)—is that the dynamical strong scale strictly exceeds the Hubble rate, $\Lambda(\bar\phi) > H$. This threshold can be understood through two complementary physical pictures: causality and thermal dynamics. 

Geometrically, the characteristic size of a hadronic bound state is $L_{\rm QCD} \sim \Lambda^{-1}$. For quarks to bind into hadrons, they must be able to causally communicate. However, the cosmological event horizon is given by $R_H \sim H^{-1}$. If the Hubble scale exceeds the confinement scale ($H \gtrsim \Lambda$), the spacetime expands so rapidly that the cosmological horizon is smaller than a hadron. Quarks are pulled apart by the expansion of the universe faster than they can exchange gluons, rendering confinement impossible. 

Equivalently, the rapid expansion of de Sitter space manifests to any spectator field as an ambient thermal bath with a Gibbons-Hawking temperature $T_{\rm GH} = H/(2\pi)$. In finite-temperature quantum chromodynamics, the vacuum undergoes a deconfinement phase transition when the temperature exceeds the critical scale $T_c \sim \Lambda$. If $T_{\rm GH} \gtrsim \Lambda(\bar\phi)$, the de Sitter background is effectively ``too hot'' to support a confined phase. In this high-temperature regime, the non-perturbative QCD instanton effects—which are responsible for tilting the Peccei-Quinn potential and generating the axion mass—are exponentially suppressed. The thermal background effectively melts the axion potential away. 

Therefore, the axion only acquires a heavy mass that can be capable of suppressing isocurvature perturbations for the early modes ($k < k_{\rm light}$) where $\Lambda(\bar\phi) > H$. As the inflaton rolls towards the end of inflation, $\Lambda(\bar\phi)$ inevitably drops below $H$. At this critical moment (corresponding to the scale $k_{\rm light}$), quarks lose causal contact, QCD deconfines, and the axion transitions into a light spectator field, allowing the late-time modes ($k > k_{\rm light}$) to freeze out with a standard unsuppressed de Sitter spectrum.

 The mass in our case doesn't change instantaneously, meaning that we cannot use Eqs.~\eqref{eq:Pthetasplit} for $k\simeq k_{\rm light}$ (see Ref.~\cite{Chakraborty:2025lyp}).
Furthermore, the axion mass during inflation and  before the transition  is time-dependent in our model, as it is given by the QCD scale in Eq.~\eqref{eq:LambdaQCD}, which depends on the rolling inflaton field $\phi$.
 However, the rate of change is small enough that it can be used as constant, as long as it is abofe $3H/2$, at least in order to build intuition. The mass is $m_a\simeq \Lambda^2/f_a$ and thus the rate of change compared to the Hubble time will be (the numbers will become apparent later in the text)
 \beq
\left |{\dot m_a\over m_a H}\right | =\left | {2\dot \Lambda\over \Lambda H}  \right | \simeq 22\sqrt{2\epsilon} {M_{\rm {Pl}}\over M} = {\cal O} (0.1)<1 \, .
 \eeq
 Although the transition between the heavy and light regimes is not instantaneous, our results --the suppression of isocurvature modes at long wavelengths and the deSitter excitations at short wavelengths-- are insensitive to the precise time dependence of the axion mass, as long as the transition occurs over a timescale shorter than the total duration of inflation. Therefore, the above discussion is sufficient to capture the essential features of our model, in particular how the existence of a super-Hubble mass suppresses isocurvature modes. Even if the condition $m_a > 3H/2$ does not hold throughout the entirety of inflation (as is typically the case in our model), the suppression of isocurvature perturbations still occurs during the period after horizon crossing for which the condition $m_a > 3H/2$ is satisfied.

We can use the approximate broken power-law spectrum to compute the variance of the axion field at the end of inflation, $\sigma_\theta^2 \equiv \langle \theta^2\rangle$. When the Hubble parameter drops below the axion mass ($3H \lesssim m_a$), the field begins to coherently oscillate. This oscillating condensate behaves as a gas of non-relativistic cold dark matter particles, with an initial number density $n_a \simeq \frac{1}{2} m_a f_a^2 \sigma_\theta^2$. Because the large-scale spatial average vanishes ($\langle \theta \rangle = 0$), the accumulated variance $\sigma_\theta^2$ completely dictates the effective misalignment angle in each Hubble patch, acting as the seed for the cosmological dark matter abundance. Given a power spectrum $P_\theta(k)$, the variance is
\beq
\langle \theta^2\rangle  = \int{ d^3k \over (2\pi)^3} P_\theta (k) = \int_{k_{\rm {min}}}^{k_{\rm {max}}} {dk\over 2\pi^2}k^2 P_\theta (k)  
\, ,
\eeq
where we used $k\in (k_{\rm {min}} ,k_{\rm {max}})$ instead of $k\in(0,\infty)$ to avoid IR and UV divergences. We could also include a smoothing function to coarse grain the spectrum. Using the broken power-law spectrum of Eq.~\eqref{eq:Pthetasplit} we get
\beq
\langle \theta^2\rangle = 
\int_{0}^{k_{\rm light}} {dk\over 2\pi^2}k^2 {H^2\over 2f_a^2} {1\over k_*^3}
+ 
\int_{k_*}^{k_{\rm {max}}} {dk\over 2\pi^2}k^2 P{H^2\over 2f_a^2} {1\over k^3} 
\, ,
\eeq
where we chose $k_{\rm {min}} =0$ as the spectrum is suppressed for small wavenumbers (large wavelengths) and thus no IR divergences arise. By performing the  integrals we get
\beq
\sigma_\theta^2\equiv \langle \theta^2\rangle = 
{H^2\over 4\pi^2 f_a^2}\left [ {1\over 3} 
+ \ln \left ( {k_{\rm {max}}\over k_{\rm light}} \right)
\right ]\simeq {H^2\over 4\pi^2 f_a^2} \Delta N \, .
\label{eq:thetavarianceuk}
\eeq
Here and throughout this work, we measure $N$ as the number of $e$-folds before the end of inflation, with $N=0$ at the end of inflation. In this convention, $\Delta N$ is always positive and denotes the duration between a given event and the end of inflation.
In Eq.~\eqref{eq:thetavarianceuk}  $\ln (k_{\rm {max}}/ k_{\rm light})=\Delta N$ corresponds to the number of $e$-folds between the transition to a light axion and the end of inflation, hence $\Delta N= N_{\rm dec}$. For simplicity we took $\Delta N\gg 1/3$. We will use the last expression $\sigma_\theta^2 = (H^2/4\pi^2f_a^2) N_{\rm dec}$ for the remainder of this work.

Overall, for $m_a > 3H/2$, the axion is driven close to the minimum  of its potential and its fluctuations are  suppressed compared to the massless case. This can be viewed either in the mode-function analysis or in the stochastic formalism (see Appendix~\ref{app:stochastic}).
A heavy axion  during inflation efficiently suppresses axion isocurvature perturbations. This allows compatibility between high-scale inflation and axion dark matter, provided the axion mass relaxes to its standard small value after inflation. Such a mechanism is therefore a natural   solution to the axion isocurvature problem. During the later parts of inflation, after the CMB modes have exited the horizon, the axion mass will inevitably drop below the Hubble scale (this is shown in Figure~\ref{fig:scalesvsN} and explained in detail in Section~\ref{sec:results}).  Then,  de Sitter fluctuations will excite axion modes with a variance which grows as a function of the number of $e$-folds and is given in Eq.~\eqref{eq:thetavarianceuk}.

In summary, a sufficiently large axion mass during inflation suppresses isocurvature perturbations, while a subsequent transition to a light axion allows for the generation of de Sitter fluctuations, which --as we will see later-- are relevant for dark matter production.

\subsection{Reheating channels}
\label{sec:reheating}

 After inflation, the inflaton will oscillate around the minimum of its potential. There is an important subtlety here. In the plateau models that we will consider, the oscillation frequency of the inflaton after inflation is typically much larger than the QCD scale, as will be demonstrated in Figure~\ref{fig:scalesvsN}.
 The typical QCD hadronization time-scale is $\tau_{\rm QCD}\sim \Lambda^{-1}$ and the typical time-scale of reheating is the inflaton oscillation frequency $\tau_{\rm reh}\sim m_\phi^{-1}$. For the models we consider $\Lambda\ll m_\phi$ and thus $\tau_{\rm reh}\ll \tau_{\rm QCD}$.
 We expect that QCD dynamics cannot follow the very short time-scales of inflaton variation and therefore we  use the time-averaged field $\langle \phi\rangle_t$ in Eq.~\eqref{eq:LambdaQCD} to describe the effective QCD scale after inflation, leading to
 \beq
\Lambda(\bar\phi) \to \Lambda(\langle \phi\rangle_t) \simeq \Lambda(0) = \Lambda_0,
 \eeq
 since $\langle \phi\rangle_t\approx 0$ during reheating. A more detailed investigation of this subtlety has both  theoretical and phenomenological interest and is left for future work.
 
\subsubsection{Minimal Reheating: inflaton decay to gluons}
 
 The tree-level perturbative decay rate of inflaton into gluons, $\phi\to gg$, is given by:
 \begin{equation}
     \Gamma_{\phi\to gg} = {m_\phi^3\over 8 \pi M^2},
 \end{equation}
where  $m_\phi$ is the inflaton mass and $M$ controls the inflaton-gluon coupling as defined in Eq.~\eqref{eq:interactionterm}.
Assuming that at the end of reheating, the resulting radiation thermalizes quickly, the energy density of radiation (which is the total energy density of the universe), will be related to its temperature through
\beq
\label{eq:rhorvsT}
\rho_r = {\pi^2\over 30 g_*}T^4 \, ,
\eeq
where the total (radiation) energy density is related to the Hubble scale through the Friedman equation $\rho = 3 H^2 M_{\rm Pl}^2$. 
Reheating is considered complete when  $H_{\rm {reh}} \simeq \Gamma_{\phi\to gg} $, allowing us to compute the reheat temperature by combining Eq.~\eqref{eq:rhorvsT} with the Friedmann equation, leading to
\begin{equation}
    T_{{\rm reh},gg}\simeq
   \left ( {90\over \pi^2 g_*} \right )^{1/4} 
    \sqrt{ \Gamma_{\phi\to gg} M_{\rm {Pl}}}
    \simeq 0.1 {m_\phi\over M_*} \sqrt{m_\phi M_{\rm Pl}} \, .
\label{eq:Trehgg}
\end{equation}
In the last step we used the effective number of relativistic degrees of freedom in the SM above the EW scale, $g_*=106.75$. 
The interaction term 
$\phi G_{\mu\nu}G^{\mu\nu}$
 provides a natural reheating channel, allowing the inflaton to decay into pairs of gluons. These gluons subsequently produce quarks and antiquarks through scattering processes, and these in turn generate the rest of the Standard Model (SM) particles, quickly leading to a thermal plasma. This process occurs irrespective of whether hadronization takes place immediately, since energetic gluons at high temperatures efficiently populate the full SM particle content. Therefore, the coupling 
$\phi G_{\mu\nu}G^{\mu\nu}$
 serves as an “economical” model-building ingredient, simultaneously responsible for raising the QCD scale during inflation and reheating the universe afterward.\footnote{Due to the bosonic nature of gluons, an oscillating infaton condensate can parametrically excite gluon modes, thereby accelerating reheating through preheating. We show in Appendix~\ref{app:preheat} that parametric resonance in our model is inefficient, and thus perturbative reheating is the only channel for the infaton to transfer the entirety of its energy into gluons and through them into a SM plasma. This justifies the use of Eq.~\eqref{eq:Trehgg}.}

\subsubsection{Maximum Reheat Temperature due to Instantaneous Reheating}
Moving away from the most minimal model, where the inflaton couples solely  to gluons, we can  introduce couplings between the inflaton and other SM (or intermediary) particles, which will  be important as long as it leads to a larger reheat temperature than Eq.~\eqref{eq:Trehgg}.  
Assuming that reheating is instantaneous, the total energy density of the universe that turns into radiation at the end of inflation is $\rho_r = 3 M_{\rm Pl}^2 H_{\rm end}^2$, where $H_{\rm end}$ is the Hubble scale at the end of inflation. Inserting this into Eq.~\eqref{eq:rhorvsT} we compute
the maximal reheat temperature, corresponding to instantaneous reheating
\beq
\label{eq:Trehmax}
T_{\rm reh,max}\simeq 
\left (
{90\over \pi^2 g_*}
\right )^{1/4}
\sqrt{H_{\rm {end}} M_{\rm Pl}}
\simeq 0.54 \sqrt{H_{\rm {end}} M_{\rm Pl}} \, .
\eeq 
This temperature  is significantly larger than the  reheat temperature from perturbative reheating into gluons (given all the other constraints). It has been shown that instantaneous reheating can be achieved through preheating via very efficient parametric or tachyonic resonance (e.g.~\cite{Sfakianakis:2018lzf, Adshead:2015pva, Cui:2021are}).

\subsubsection{An example beyond Minimal Reheating:  Inflaton decay to heavy right handed neutrinos}
\label{sec:reheattonu}
Above we have studied the minimal case in which reheating takes place only via the inflaton/gluon coupling that we used to drive the early QCD confined phase as a solution to the isocurvature problem.   Now we turn to an extension beyond this minimal model in which reheating takes place to additional particles.  Reheating to bosons would require an analysis of nonperturbative reheating including parametric resonance. Here, instead,  
we will invoke a simple Yukawa coupling to heavy right-handed neutrinos\footnote{{
Since the inflaton is a SM singlet, Yukawa coupling to SM fermions is forbidden. The lowest allowed operator necessarily includes the Higgs field to couple left and right-handed Weyl fermions, ${\cal L}\subset c (\phi/\Lambda) \bar L_L H e_R$. If the electroweak symmetry is broken, as it typically is during inflation, the Higgs VEV will make this interaction term look like a Yukawa coupling of the inflaton to Dirac fermions. After inflation, the inflaton will start decaying following the decay rate of Eq.~\eqref{eq:decayrate}. Once a thermal bath appears during reheating, hot enough to restore electroweak symmetry, SM fermions stop behaving like Dirac fermions and inflaton decay will need to include the two fermions and the Higgs, making this a three body decay process. We do not pursue this decay channel further.}}, of the form
\beq
\label{eq:yukawa}
{\cal L}_{\phi f\bar f} = \sum_i y_i\phi f_i \bar f_i,
\eeq
We can choose the Yukawa couplings of the inflaton to each particle species to be different, thereby allowing the inflaton to decay preferentially. For simplicity, we 
choose a universal Yukawa coupling $y\equiv y_i$, leading to the decay rate of the inflaton to heavy right handed neutrinos:
\beq 
\label{eq:decayrate}
\Gamma_{\phi \to f\bar f} = {{\cal N}_f\over 8\pi} y^2 m_\phi  \to 0.12 \, y^2 m_\phi\, ,
\eeq
where ${\cal N}_f$ is the number of heavy neutrinos that  couple to the inflaton, which we also took to be much lighter than the inflaton at the time of reheating. In the last step we assumed three Dirac neutrinos. 
The reheat temperature\footnote{{We assume that the right handed neutrinos can decay or annihilate into SM leptons at a fast enough rate to produce a thermal plasma during reheating. For example, an interaction term between right handed neutrinos and the SM of the form $\lambda N_R H L $ can allow for a fast (faster than Hubble) annihilation rate between right-handed neutrinos for $\lambda\gtrsim 0.01$.}} becomes $T_{\rm {reh},y} \simeq 0.54 \sqrt{\Gamma_{\phi \to f\bar f} M_{\rm {Pl}}}$ .

Considering more or fewer right-handed neutrinos will change the pre-factor in Eq.~\eqref{eq:decayrate}. Furthermore, heavy Majorana neutrinos are also a viable decay channel, which we will not pursue in detail here. These details (number and nature of neutrinos)  affect how easy or hard it is to achieve certain  outcomes, like fast or instantaneous reheating. 

An important restriction on inflationary model-building arising from the introduction of Yukawa coupling to fermions comes in the form of loop corrections, that can alter (or dominate over) the tree-level inflationary potential~\cite{Ellis:2025bzi}. The coupling of Eq.~\eqref{eq:yukawa} induces corrections to the inflaton potential of the form~\cite{Markkanen:2018bfx} 
\beq
\Delta V_y = - {{n}_f\over 64\pi^2}
y^4\phi^4 \left [
\log \left (
{y^2\phi^2\over\mu_{\rm {ren}}^2}
\right) -{3\over 2}
\right
],
\eeq
where  $n_f$ is the total number of fermionic degrees of freedom contributing in the loop (for example, each Dirac fermion contributes four fermionic degrees of freedom) and $\mu_{\rm {ren}}$ is the renormalization scale. We can choose the renormalization scale, in order to minimize the logarithm, $\mu_{\rm {ren}}\simeq y\phi$. 
Since we do not aim to provide an RG-improved inflationary potential, but simply to restrict the Yukawa couplings so that $\Delta V_y\ll V_{\rm{infl}}$, we will use the simplified formula, neglecting the logarithmic dependence, as
\beq
\Delta V_{y,{\rm simpl}} \simeq {{n}_f\over 64\pi^2}
y^4\phi^4 
= {{\cal{N}}_f\over 16\pi^2}
y^4\phi^4  \,, 
\label{eq:deltaVsimple}
\eeq
where  $n_f=4{{\cal N}}_f$ if we consider Dirac fermions coupled to the inflaton. As we noted for the decay rate of Eq.~\eqref{eq:decayrate}, coupling the inflaton only to {three right-handed neutrinos} leads to ${\cal N}_f=3$.  Majorana neutrinos ``count" differently, but we will not consider this here. 
We will return to this correction, after discussing  the inflationary potential in detail in Section~\ref{sec:realization}.

\subsubsection{Avoiding restoration of PQ Symmetry during Reheating}

Before concluding the section on reheating, let us make the important distinction between two characteristic temperatures. The reheat temperature $T_{\rm reh}$ is the temperature of the thermalized plasma at the end of reheating, when the inflaton energy density has been fully transferred to radiation and standard radiation domination commences, corresponding to $H \sim \Gamma_\phi$.  The maximum temperature $T_{\rm max}$, on the other hand, is the highest temperature reached during the reheating process, occurring shortly after the onset of inflaton oscillations. 
After this peak, the plasma cools as $T \propto a^{-3/8}$ until it reaches $T_{\rm reh}$ at the end of reheating when radiation domination begins. Crucially, $T_{\rm max} \ge T_{\rm reh}$, with the hierarchy given by~\cite{Giudice:2000ex}
\beq
T_{\rm max}= \left ({3\over 8}\right )^{2/5} \left (
{40\over \pi^2}\right )^{1/8} 
{g_*^{1/8}(T_{\rm reh}) \over g_*^{1/8}(T_{\rm max})  } M_{\rm Pl}^{1/4} H_{\rm end}^{1/4} T_{\rm reh}^{1/2}
\simeq 0.45  (M_{\rm Pl} H_{\rm end} T_{\rm reh}^{2})^{1/4} \, ,
\eeq
where in the last step we took $g_*(T_{\rm reh})=g_*(T_{\rm max}) =106.75$.
The equality holds only for instantaneous reheating and the hierarchy between the two temperatures grows as $T_{\rm reh}$ becomes lower.

Throughout this discussion we assume efficient and instantaneous thermalization of the inflaton decay products, so that a well-defined plasma temperature can be assigned at all times during reheating. The distinction between $T_{\rm max}$ and $T_{\rm reh}$ is phenomenologically important, as the production of heavy relics such as superheavy dark matter is sensitive to the highest temperature reached, not merely $T_{\rm reh}$~\cite{Chung:1998rq}. For us, the highest temperature reached after inflation is a key quantity in determining whether or not the PQ symmetry is restored in the early universe. Our calculations require PQ symmetry to remain broken after inflation.  However, PQ symmetry could be restored if $T_{\rm max}$ were high enough. A necessary condition for PQ symmetry restoration would be $T_{\rm max} \gtrsim f_a$; further  the PQ sector would have to thermalize with the plasma, requiring $\Gamma_{\rm PQ}/H \gtrsim 1$ at $T \sim f_a$.  The thermalization condition $\Gamma_{\rm PQ}/H \gtrsim 1$ depends sensitively on the specific realization of the PQ sector\footnote{This model dependence becomes even more pronounced in supersymmetric realizations, where one must track the full PQ scalar sector rather than only the axion angular mode.} and its couplings to the plasma. For simplicity, we adopt the necessary—but not sufficient—condition $T_{\rm max} \gtrsim f_a$ as our criterion for PQ symmetry restoration. This constitutes a conservative and model-independent approach, as PQ symmetry restoration may not occur even when $T_{\rm max} \gtrsim f_a$ if the interaction rate is too small. Since the validity of our calculations requires PQ symmetry to remain broken after inflation -- otherwise one must track axion production from cosmic strings and domain walls -- we exclude regions of parameter space where $T_{\rm max} \gtrsim f_a$.

In summary, reheating determines both the thermal history of the Universe and the possibility of PQ symmetry restoration, with the maximal temperature $T_{\max}$ providing a conservative criterion for assessing whether the PQ symmetry remains broken after inflation.

\subsection{DM abundance}
\label{DM}

The axion DM abundance today can receive contributions from three sources: a residual homogeneous displacement $\bar\theta$, de Sitter quantum fluctuations with standard deviation $\sigma_{\theta,{\rm dS}}\sim (H/2\pi f_a)\sqrt{N_{\rm dec}}$, and thermal fluctuations generated during reheating. Here $\bar\theta$ denotes the coarse-grained axion angle in our Hubble patch after the heavy-axion phase, which later sets the homogeneous misalignment contribution when the axion begins its standard QCD oscillations. In the mechanism considered here, the axion is heavy during the early stages of inflation and is dynamically driven toward the minimum of its potential, so we take $\bar\theta=0$. The abundance is therefore controlled by the stochastic de Sitter and thermal contributions.

Having analyzed the de Sitter contribution in Section~\ref{sec:deSitter}, we now estimate the stochastic displacement that the axion may receive from the plasma produced during reheating. After inflation, the inflaton oscillates about the minimum of its potential. In the parameter range of interest, these oscillations are much faster than the QCD time scale, so the QCD sector effectively responds to the time-averaged inflaton background, for which $\langle\phi\rangle_t\simeq 0$. The confinement scale relevant during reheating is therefore approximately the ordinary QCD scale, $\Lambda_{\rm QCD}\sim 400\,{\rm MeV}$.

A temperature is not defined immediately at the onset of reheating, when the energy density is still dominated by the coherent inflaton condensate. Once a population of decay products is produced and approximately thermalized, however, the plasma can be assigned an effective temperature. In the regimes considered below this temperature, whether characterized by $T_{\rm reh}$ or by the transient maximum temperature $T_{\max}$, is far above $\Lambda_{\rm QCD}$. The finite-temperature QCD axion potential is then strongly suppressed, and the axion is effectively light.

If the axion, or the PQ sector more generally, interacts with this plasma, the long-wavelength axion mode can be treated as an open system. Here this simply means that we keep only the Hubble-patch averaged axion field, while the shorter-wavelength plasma degrees of freedom are treated as an environment. Interactions with this environment produce both dissipation and stochastic noise. After integrating out the bath degrees of freedom, the coarse-grained axion field obeys a Langevin equation of the form
\begin{equation}
\ddot a + (3H+\Upsilon)\dot a \simeq \xi(t),
\end{equation}
where $\Upsilon(T)$ is a dissipation coefficient and $\xi$ is a stochastic noise term. The fluctuation--dissipation relation gives, parametrically,
\begin{equation}
\langle \xi(t)\xi(t')\rangle 
\simeq \frac{2\Upsilon T}{V}\,\delta(t-t'),
\end{equation}
with $V\sim H^{-3}$ the coarse-graining volume of one Hubble patch.

In the light-field regime, $m_a(T)\ll H$, the curvature of the axion potential can be neglected over a Hubble time. If the coarse-grained mode is also friction dominated, so that the acceleration term is subleading compared with $(3H+\Upsilon)\dot a$, the stochastic displacement over a time interval $\Delta t\sim H^{-1}$ is estimated as
\begin{equation}
\langle (\Delta a)^2\rangle 
\sim \frac{2\Upsilon}{(3H+\Upsilon)^2}\,H^2 T.
\end{equation}
For efficient but perturbative coupling to the plasma, $\Upsilon\sim H$, the prefactor is order unity and the variance of the Hubble-patch averaged axion field scales as
\begin{equation}
\langle a^2\rangle_{k\lesssim aH} \sim H T,
\qquad
\sigma_{\theta,{\rm th}}^2 \sim \frac{H T}{f_a^2}.
\end{equation}
This estimate refers to the coherent axion displacement within one Hubble patch. Shorter-wavelength thermal fluctuations average out and do not contribute directly to the effective misalignment angle.

In applying this estimate, we take $(H_{\rm reh},T_{\rm reh})$ as a conservative reference point, since this corresponds to the onset of radiation domination when the thermal bath is long-lived. During perturbative reheating the plasma can temporarily reach a higher maximum temperature, $T_{\max}>T_{\rm reh}$, while the universe is still inflaton dominated. If axion interactions are efficient already at this earlier stage, $\Gamma_a(T_{\max})\gtrsim H(T_{\max})$, the stochastic forcing may instead be controlled  by $(H(T_{\max}),T_{\max})$\footnote{If axions thermalize with the other particles in the bath in the early Universe, a thermal population of axions will be produced and eventually freezes out. For axion masses above $\sim{\rm eV}$, the resulting thermal axion population constitutes a hot DM component, which is tightly constrained to be a small fraction of the total matter energy density~\cite{Hannestad:2010yi,Cadamuro:2010cz,Archidiacono:2013cha,DiValentino:2015wba}. For lighter axions, the thermal population behaves as dark radiation. In the case of thermalization at temperatures above the electroweak scale, which is relevant for the present study, the contribution of thermal axions to dark radiation is not currently constrained by experiments~\cite{Baumann:2016wac}.}.
However, since the $T_{\max}$ phase is short-lived and subsequent entropy
production dilutes early excitations, $T_{\rm reh}$ provides a conservative
and model-independent reference scale.

The dissipation coefficient $\Upsilon$ arises from the same underlying
interactions that determine the microscopic axion interaction rate
$\Gamma_a$ in the plasma; parametrically one expects
$\Upsilon \sim \Gamma_a$.
A detailed determination of the dissipation coefficient $\Upsilon$ and the corresponding axion interaction rate depends on the specific realization of the PQ sector and its couplings to the thermal bath; here we adopt a parametric treatment that captures the relevant scaling without committing to a specific model.
We therefore parametrize the thermal contribution as
\begin{equation}
\sigma_{\theta,{\rm th}}^2
= \epsilon_{\rm th}\,\frac{H_*T_*}{f_a^2},
\qquad
\epsilon_{\rm th} \sim \min\!\left(1,\frac{\Gamma_a}{H}\right),
\end{equation}
Here $\epsilon_{\mathrm{th}} \leq 1$ parametrizes the efficiency of thermalization and the duration of the interaction between the axion and the plasma; it effectively controls the strength of thermal fluctuations and is treated as a free parameter in our analysis.
In practice, when using the above equation, we take $T_* = T_{\rm reh}$   as a conservative reference scale, while $T_{\max}$ represents the highest temperature reached during reheating and may lead to stronger thermal effects if interactions are efficient at earlier times.

As a complementary upper envelope, one may assume that thermal interactions randomize the axion field up to the full thermal scale, $\delta a\sim T$, giving
\begin{equation}
\sigma_{\theta,{\rm th}}^2 \lesssim \frac{T^2}{f_a^2}.
\end{equation}
This should be regarded as a maximal estimate rather than the generic Hubble-patch averaged result. In what follows we use both estimates to bracket the possible size of thermal effects: the conservative coarse-grained estimate $\sigma_{\theta,{\rm th}}^2\sim \epsilon_{\rm th}HT/f_a^2$ and the upper-envelope estimate~\cite{Dvali:1995ce,Koutsangelas:2022lte} $\sigma_{\theta,{\rm th}}^2\sim T^2/f_a^2$.

Let us now consider axion production from thermal fluctuations in addition to the corresponding one from de Sitter fluctuations.
Since the two  contributions are stochastic and independent, their variances add, leading to
\beq
\label{eq:variancefull}
\sigma_\theta^2 = \sigma_{\theta,dS}^2 + \sigma_{\theta,{\rm{thermal}}}^2 \simeq {H^2\over 4\pi^2 f_a^2} N_{\rm dec} +\frac{T_{\rm equiv}^2}{f_a^2} \, .
\eeq
The DM abundance is proportional to $\sigma_\theta^2$, given in Eq.~\eqref{eq:variancefull}.
For notational simplicity, we defined an ``equivalence" temperature
\beq
\label{eq:Tequiv}
T_{\rm equiv}^2 \equiv \begin{cases}
T_{\rm reh}^2 & \text{upper~bound}  \\
\epsilon_{\rm th} \, H_{\rm reh} \, T_{\rm reh} \simeq 3.4 \epsilon_{\rm th} { T_{\rm reh}^3\over M_{\rm Pl}}& \text{Hubble-averaged~estimate} 
\end{cases}
\eeq
In the above equation $~\epsilon_{\rm th}\le    1 $ and in the last step the equation for the Hubble scale at reheating was used
\beq
H_{\rm reh} = \sqrt{\pi^2 g_*(T_{\rm reh})\over 90} {T_{\rm reh}^2\over M_{\rm Pl}}\, ,
\eeq
with $g_*\simeq 106.75$.

It is useful to distinguish between different regimes depending on which contribution dominates the axion variance. We consider three limiting cases:
\begin{enumerate}[label=(\roman*)]
    \item  \emph{De Sitter--dominated regime}, when $\sigma^2_{\theta,\mathrm{dS}} \gg \sigma^2_{\theta,\mathrm{th}}$, corresponding to sufficiently low reheating temperature or inefficient thermalization ($\epsilon_{\mathrm{th}} \ll 1$);
\item \emph{thermal-dominated regime}, when $\sigma^2_{\theta,\mathrm{th}} \gg \sigma^2_{\theta,\mathrm{dS}}$, typically realized for high reheating temperatures and efficient interactions with the thermal bath ($\epsilon_{\mathrm{th}} \sim 1$);
and 
\item \emph{mixed regime}, when the two contributions are comparable.
\end{enumerate}
In practice, we do not treat these regimes separately, but instead use the full expression in Eq.~\eqref{eq:variancefull}, allowing the system to determine dynamically which contribution dominates across parameter space.
The parameter $\epsilon_{\mathrm{th}}$ controls the efficiency of thermal fluctuations and encodes our uncertainty regarding the strength and duration of the interaction between the axion and the plasma.
In the following, we consider representative values of $\epsilon_{\mathrm{th}}$ to bracket the possible thermal contributions.

Up to the usual ${\cal O}(1)$ uncertainties associated with the temperature dependence of the QCD axion mass and anharmonic corrections, the axion abundance scales as
\begin{equation}
    {\Omega_a\over \Omega_{\rm DM}}\sim \left(
{f_a\over 10^{12} \, {\rm {GeV}}}
    \right )^{7/6}
  \sigma_\theta^2
  \simeq \left (  {f_a\over 10^{12} \, {\rm {GeV}}}  \right )^{7/6}
{T_{\rm {equiv}}^2 + {N_{\rm dec}\over 4\pi^2}H^2 \over f_a^2} \, ,
\label{eq:DMabundace}
\end{equation}
where we have used $\sigma_\theta^2$ from Eq.~\eqref{eq:variancefull}, which 
depends on the details of the thermal history and axion interactions strength with the plasma.

 Unless otherwise stated, we assume that the QCD axion accounts for the entirety of dark matter.
We can thus invert this equation to compute the axion decay constant $f_a$ in terms of the reheat temperature, Hubble scale and time (in $e$-folds) of the decoupling transition during inflation. This leads to
\beq
\label{eq:faDM}
{f_a\over M_{\rm Pl}} = \left ( 2.8\times 10^7 \,  {T_{\rm {equiv}}^2 + { N_{\rm dec}\over 4\pi^2}H^2 \over M_{\rm {Pl}}^2} \right)^{6/5} \, .
\eeq
This expression should be interpreted as the value of $f_a$ required for the QCD axion to constitute the entirety of the observed dark matter for a given inflationary scale, reheating history, and duration of the light-axion phase.

In summary, the axion dark matter abundance receives contributions from both de Sitter and thermal fluctuations, with their relative importance determined dynamically by Eq.~\eqref{eq:variancefull} and controlled by the reheating temperature and the efficiency parameter $\epsilon_{\mathrm{th}}$.

\subsection{Sequence of scales}
\label{sec:scales}

Before attempting a concrete realization, let us list the necessary conditions that are needed for this model to be self-consistent and effective at suppressing axion isocurvature modes.
\begin{itemize}

\item The energy density of the quark-gluon sector must be subdominant to the energy density of the inflaton, $\Lambda^4 <V$, in order not to dominate the inflationary dynamics. At the same time, during the CMB-relevant epoch of inflation, the QCD sector must be confined, equivalently the QCD scale must be higher than the Hubble scale $\Lambda>H$.  This is necessary for the axion potential to arise.\footnote{The requirement for early QCD confinement during inflation is that the dynamical scale exceeds the Hubble rate, $\Lambda > H$. Physically, this can be understood through causality: the characteristic size of a hadronic bound state, $\Lambda^{-1}$, must fit within the cosmological event horizon, $H^{-1}$. If $H > \Lambda$, the spacetime expands so quickly that quarks cannot effectively communicate to be confined. Equivalently, in the thermal picture of de Sitter space, this regime corresponds to the Gibbons-Hawking temperature $T_{\rm GH} = H/(2\pi)$ exceeding the confinement scale, which suppresses the non-perturbative instanton effects responsible for the axion mass. Thus, as the inflaton rolls and $\Lambda(\phi)$ drops below $H$, a deconfinement transition is triggered, immediately rendering the axion light. }
Altogether ${H<\Lambda<V^{1/4}}$. 

\medskip

\item The axion mass must be larger than (or at least comparable to) the Hubble scale during inflation, in order to suppress isocurvature fluctuations during the CMB-relevant window, more precisely ${m_a>3H/2}$, as shown in Section~\ref{sec:deSitter}, where  $m_a = \Lambda^2/f_a$.

\item Expanding the cosine potential of the axion around its minimum and examining the quartic term, the perturbative unitarity limit requires ${\Lambda<f_a}$ (technically there is a factor of $(8\pi)^{1/4}\simeq 2.2$ in front of $f_a$, which we neglect. This condition can be used to also write $m_a = \Lambda^2/f_a<\Lambda$.

\item Given our discussion at the end of Section~\ref{sec:reheating}, we require $T_{\rm max}<f_a$ as a conservative criterion for the non-restoration of the PQ symmetry. 

\item Last but not least, we wish to keep the theory under control, thus require ${M<\Mp}$.
\end{itemize}
We can summarize all conditions on the different mass- and energy-scales as
\beq
\label{eq:inequalities1}
\boxed{3H/2<m_a <\Lambda<\{f_a , \, V^{1/4} \} ~ ,\quad M<\Mp \,  \quad T_{\rm max}<f_a}\, .
\eeq

The above conditions ensure that the theory is perturbative and self-consistent, that the inflaton sector dominates the inflationary dynamics and that during the time of interest the axion is heavier than the Hubble scale and thus not excited, thereby suppressing isocurvature perturbations on cosmologically observable scales.

\section{A Concrete Inflationary Realization as an example: $\alpha$-attractor inflation}
\label{sec:realization}

In this section we   review the dynamics of $\alpha$-attractor inflationary models and use them as an example of a concrete realization of our scenario. We  also examine how loop corrections due to fermion couplings, as well as corrections to the inflationary potential from the presence of a confined QCD sector during inflation, lead to constraints on the model parameters.

Let us start with a quick overview of simple $\alpha$-attractor models and their inflationary dynamics. The two prototypical models of $\alpha$-attractors are the E- and T-models, 
\begin{eqnarray}
    V_E &=& V_0 \left (
1-e^{-\sqrt{2\over 3\alpha}\phi}
    \right ) ^2 \, ,
    \\
V_T &=& V_0 \tanh^2\left (
{\phi\over \sqrt{6\alpha}}
\right),
\end{eqnarray}
where $V_T$ is symmetric about $\phi = 0$.
We do not wish to provide a complete history or description of $\alpha$-attractors (see e.g. Refs.~\cite{Carrasco:2015uma, Carrasco:2015pla,Galante:2014ifa, Kallosh:2013yoa, Kallosh:2013hoa, Iarygina:2018kee}).
The parameter $\alpha$ 
 is related to the curvature of the K{\"a}hler manifold in supergravity embeddings. When derived from supergravity, $\alpha$-attractors are necessarily two field models, with one field stabilized during inflation (but active during preheating~\cite{Iarygina:2018kee, Krajewski:2018moi}). The Ricci curvature of the two-field field-space is
 ${\cal R} = -4/3\alpha$, so $\alpha$ has units of mass-squared. There may be slightly different definitions of $\alpha$-attractors, introducing ${\cal O}(1)$ factors in $\alpha$ and ${\cal R}$. We focus only on the T-model; the differences in the inflationary trajectories between the E and T models are studied in Ref.~\cite{Iarygina:2020dwe},  showing   a shift in the field between E- and T-models. An important advantage of the T-model in the present context   is that the inflaton field $\phi$ can be either positive or negative, meaning that the condition $\phi<0$, dictated by the sign of the coupling term $\phi GG$  works for inflation. We therefore choose the simplest T-model potential
 \beq
 \label{eq:potential}
V(\phi) = V_0 \tanh^2\left(
{\phi\over \sqrt{6\alpha}}
\right ) \, .
 \eeq
We will show that $V_0=\mu^2 \alpha$ with $\mu\simeq 10^{-5} \, M_{\rm {Pl}}$ reproduces the correct  spectrum of density perturbations, which is a staple of these models. Readers that are familiar with the dynamics of $\alpha$-attractors may proceed to Section~\ref{sec:results}.

 Let us look at the inflationary dynamics of this model. The Klein-Gordon equation is
 \beq
\ddot \phi + 3 H \dot\phi + V_{,\phi}=0.
 \eeq
We use the standard slow-roll approximations. Due to the flatness of the potential we take $V\simeq V_0$ in the Hubble scale, thus $H\simeq \sqrt{V_0/3} / M_{\rm {Pl}}$. We  make a similar approximation in the potential derivative and work in the regime $|\phi|/\sqrt{\alpha}\gg 1$ leading to
\beq
V_{,\phi} = \sqrt{2\over 3\alpha}
V_0\,  {\rm sech}^2 \left (
{\phi\over \sqrt{6\alpha}}
\right) \tanh{\left (
{\phi\over \sqrt{6\alpha}}
\right)}\simeq {4 \sqrt{\frac{2}{3\alpha}} 
V_0 e^{-{\sqrt{\frac{2}{3\alpha}} \phi }}}
\, ,
\eeq
where we took $\phi<0$, $\tanh(x)\simeq -1$ and $\rm{sech}(x) \simeq 2 e^{x}$ for $x\ll -1$.

Using the above approximations, along with the slow-roll condition $\ddot\phi\approx 0$ leads to
\beq
3H \dot\phi +V_{,\phi}\simeq 0,
\eeq
which can be trivially solved to give
\beq
 \dot\phi = - {V_{,\phi}\over 3H}
 \simeq \frac{4\sqrt{2}}{3}\Mp \mu \, e^{\sqrt{\frac{2}{3\alpha}}\phi},
\eeq
where we used $V_0=\alpha \mu^2$ in the last step.

We can integrate the definition of the $e$-folding number $dN = H dt$ to compute the number of $e$-folds $N$ as a function of the field value $\phi$, using the formula for $\dot\phi$ given above.
\beq
\label{eq:efoldsvsphi}
 N=\int_{t}^{t_{\rm end}} H dt = \int_{\phi}^{0} {H\over \dot\phi}d\phi = \frac{{3} \alpha}{8 \Mp^2 } e^{-\sqrt{\frac{2}{3\alpha}} \phi } \, .
\eeq
We must note a few points about Eq.~\eqref{eq:efoldsvsphi}. Firstly, we neglected the contribution from $\phi_{\rm {end}}$, the value of $\phi$ at the end of inflation. Furthermore, $|\phi|$ is larger deeper into inflation and goes towards zero as inflation comes to an end.
Therefore, $N$  has a large positive value early during inflation and goes towards zero as inflation ends. This should not be confused with  $\Delta N$ used in Eq.~\eqref{eq:thetavarianceuk}, which is defined as the number of efolds during which the axion is light. 
We can invert Eq.~\eqref{eq:efoldsvsphi} to derive the field value as a function of the $e$-folding number
\begin{equation}
\label{eq:phivsN}
    \phi = - \sqrt{3\alpha\over 2}\log \left({8M_{\rm Pl}^2   \over 3\alpha}N\right ).
\end{equation}

The first and second slow-roll parameters can be calculated straightforwardly from the potential as 
\begin{eqnarray}
\label{eq:epsilonSR}
    \epsilon &\equiv& -{\dot H\over H^2} \simeq {M_{\rm Pl}^2\over 2 }\left (
{V_{,\phi}\over V}
    \right )^2  
    \simeq
    {16 \Mp^2\over 3\alpha} e^{2{\sqrt{2\over 3\alpha}\phi}}
   \simeq  {3\over 4N^2} {\alpha\over M_{\rm Pl}^2} 
   \, ,
 \\
 \label{eq:etaSR}
 \eta_V &\equiv &
 \Mp^2 {V_{,\phi\phi}\over V} \simeq -{8\Mp^2\over 3\alpha}
 e^{{\sqrt{2\over 3\alpha}\phi}} \simeq -{1\over N} \, .
\end{eqnarray}
In both cases, $\epsilon$ and $\eta_V$, we used Eq.~\eqref{eq:efoldsvsphi} to swap $\phi$ for $N$.
The end of inflation occurs at $\epsilon=1$ which translates to $|\phi_{\rm end}| \simeq
{1\over 2}\sqrt{3\alpha\over 2} \log\left (
16\Mp^2/3\alpha
\right )= 
{\cal O}(1) \sqrt{\alpha}$.

\begin{figure}
    \centering
    \includegraphics[width=0.43\linewidth]{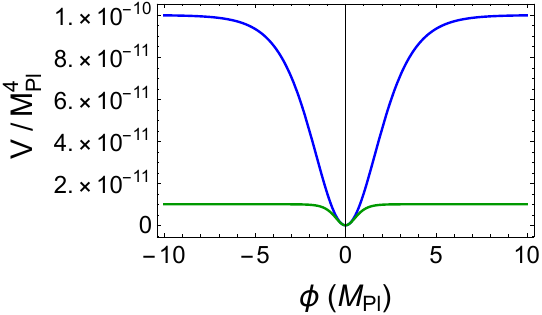}
    \includegraphics[width=0.45\linewidth]{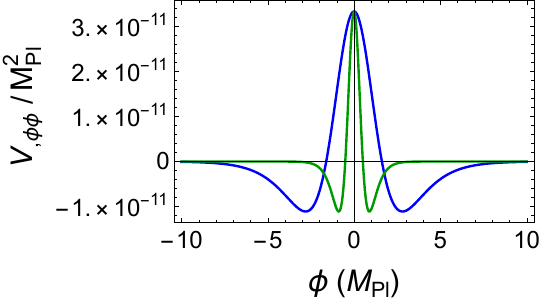}
    \caption{The potential energy (left) and effective mass-squared (right) as a function of the inflaton field, all in terms of the reduced Planck mass. The blue and green curves correspond to $\alpha/M_{\rm Pl}^2=1,0.1$ respectively. We see that the height of the plateau depends on $\alpha$, but the inflaton mass near the origin, $m_\phi^2 \equiv V_{,\phi\phi}|_{\phi=0}$ is independent of $\alpha$, as expected.}
    \label{fig:alphaattractor}
\end{figure}

The tensor-to-scalar ratio is
$
r=16\epsilon = (12/N^2) {\alpha/M_{\rm Pl}^2} 
\simeq 0.003{\alpha/M_{\rm Pl}^2}
$ for $N=60$. 
The spectral tilt is controlled by the second slow-roll parameter (since $\epsilon \ll |\eta_V|$);  $n_s = 1-6\epsilon+2\eta_V\simeq 1 -2 \eta_V \simeq 1-{2/ N}=0.9667$ for $N=60$.
Finally the scalar power spectrum amplitude is
\beq
A_s = {H^2\over 8\pi^2 M_{\rm Pl}^2 \epsilon} = {V_0\over \alpha M_{\rm Pl}^2} {N^2\over 18\pi^2} \simeq 20 {V_0\over \alpha M_{\rm Pl}^2},  
\eeq
for $N=60$. Requiring $A_s\simeq 2\times 10^{-9}$ leads to
\beq
\sqrt{V_0\over \alpha } \simeq 10^{-5}M_{\rm Pl}.
\eeq
This  makes the definition
\beq
\label{eq:V0intermsofalpha}
V_0 = \mu^2 \alpha,
\eeq
a very useful one to make for these models. 
After the end of inflation, as the field approaches the origin and reheating begins, we can expand  the hyperbolic tangent in Eq.~\eqref{eq:potential} about the origin and the quadratic term will determine the oscillation frequency of the inflaton around the origin 
\beq
V\simeq {1\over 2} {V_0\over 3\alpha} \phi^2 +...\, .
\eeq
Thus the oscillation frequency near the origin (inflaton mass) is 
\beq
m_\phi \equiv \sqrt{V_{,\phi\phi}|_{\phi=0}}
\sim \sqrt{V_0\over \alpha } \sim {\cal O}(10^{-5})M_{\rm Pl} \, ,
\eeq
independently of the value of $\alpha$.
This is a typical feature of $\alpha$-attractors. 
The height of the potential and hence the scale of inflation is set by $\alpha$ via Eqn. \eqref{eq:V0intermsofalpha},
whereas
the small field oscillation frequency (the inflaton mass) is independent of $\alpha$ and equal to the usual quadratic inflation mass. This can be seen in Figure~\ref{fig:alphaattractor}.
Thus, for perturbative reheating,  the mass of the inflaton is always the same, regardless of the Hubble scale during inflation.

\subsection{Constraints from the inflationary realization}

Each specific realization of inflation gives rise to further constraints not previously discussed.  
\subsubsection{Constraints on Yukawa Couplings in scenarios with reheating to neutrinos due to loop corrections to the potential:}
\begin{figure}
    \centering
    \includegraphics[width=0.4\linewidth]{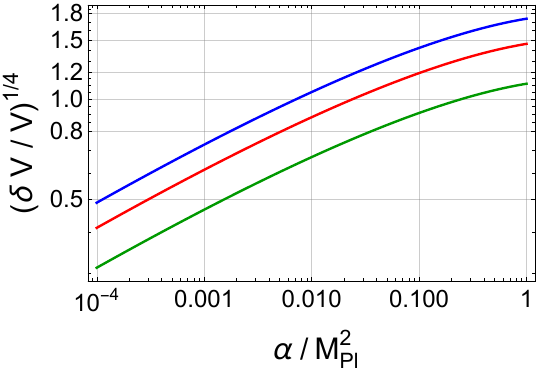}
    \includegraphics[width=0.46\linewidth]{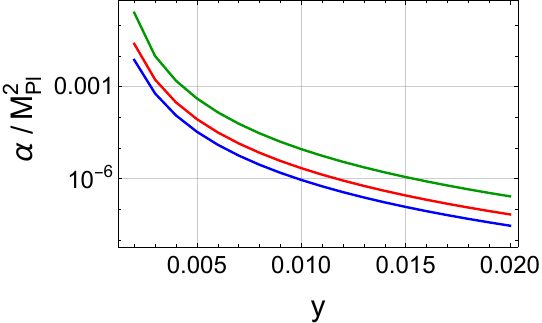}
    \\
    \includegraphics[width=0.4\linewidth]{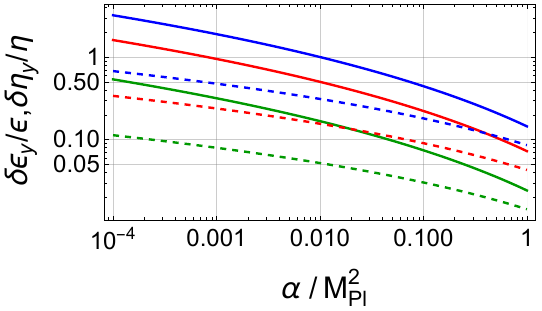}\includegraphics[width=0.46\linewidth]{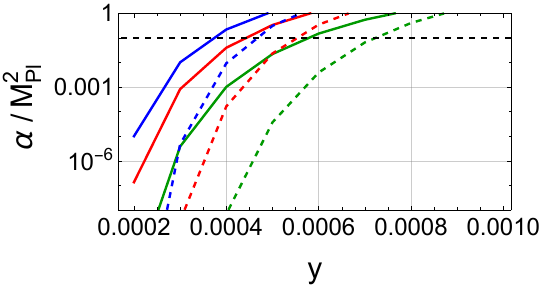}
    \caption{{\it Upper left:} The relative correction to the inflationary potential due to fermion loops, as defined in Eq.~\eqref{eq:deltaVoverVy} with $y=0.002$. The blue, red and green curves correspond to ${\cal N}_f=6,3,1$ respectively, where where ${\cal N}_f$ is the number of heavy neutrinos that  couple to the inflaton. All curves are monotonically increasing with $\alpha$, and hence for values of $\alpha$ less than the one corresponding to $\delta V/V=1$, the fermion loop corrections to the magnitude of the potential become progressively less important. 
    {\it Upper right:} The value of $\alpha$ corresponding to $\delta V/V=1$ as a function of the Yukawa coupling for ${\cal N}_f=6,3,1$ (blue, red and green respectively). For values of $\alpha$ below each curve, we will ignore fermion loop corrections to the inflationary potential. 
    \\
   {\it Lower left:} The relative correction to the slow roll parameters  due to fermion loops for $y=0.0003$. Solid and dashed curves correspond to $\epsilon$ and $\eta$ respectively. The color coding for ${\cal N}_f=6,3,2$ follows the upper panels. All curves are monotonically decreasing and hence for values of $\alpha$ above than the ones corresponding to $\delta \epsilon/\epsilon,\delta\eta/\eta=1$, the fermion loop corrections to the magnitude of the potential become progressively less important. 
    {\it Lower right:} The value of $\alpha$ corresponding to $\delta \epsilon_y/\epsilon=1$ (solid) and $\delta \eta_y/\eta=1$ (dashed). 
    }
    \label{fig:deltaVoverV}
\end{figure}

Here we discuss further constraints on the $\alpha$-attractor inflation scenario for the case of reheating via inflaton decay to heavy right handed neutrinos discussed in Section \ref{sec:reheattonu}.
We start by examining the loop corrections to the potential of Eq.~\eqref{eq:potential} by the Yukawa coupling of Eq.~\eqref{eq:yukawa}. The correction is given by Eq.~\eqref{eq:deltaVsimple} and we can write the ratio of the correction to the potential  as
\begin{equation}
\label{eq:deltaVoverVy}
 \left ({
\delta V_{y} \over V
} \right )^{1/4}  \simeq \left ({
\delta V_{y,{\rm {simpl}}} \over V
} \right )^{1/4} 
\simeq
\left ( { {{\cal{N}}_f\over 16\pi^2}
y^4\phi^4  \over V_0} \right )^{1/4}
=  
{y\over 2\sqrt{\pi\mu} }  {{\cal{N}}}_f^{1/4} {|\phi|\over \alpha^{1/4}}\, ,
\end{equation}
where we used the plateau value for the potential $V\simeq V_0= \mu^2\alpha$ and where ${\cal N}_f$ is the number of heavy neutrinos that  couple to the inflaton.
We should note that $|\phi|\propto \sqrt{\alpha}$ in the slow-roll expression of Eq.~\eqref{eq:phivsN}, if we neglect the weaker logarithmic dependence on $\alpha$, and thus $ \left (
\delta V_{y,{\rm {simpl}}} / V
 \right )^{1/4} \propto y \alpha^{1/4}{{\cal{N}}}_f^{1/4} $.
 Figure~\ref{fig:deltaVoverV} shows the value of $ \left (
\delta V_{y,{\rm {simpl}}} / V
 \right )^{1/4}$ as a function of $\alpha$ for different values of $y$ as well as the critical value of $\alpha$ for which $\delta V_y/V$=1 as we vary ${\cal N}_f$ and y.
 We will further see how requiring this ratio to be less than unity affects the available parameter space of the model in Section~\ref{sec:results}.
That being said, the criterion $\delta V_y<V$ is  necessary for guaranteeing that the loop corrections do not dominate over the potential, but not sufficient to guarantee that the loop corrections do not change the inflationary dynamics. For that we need to compute the first and second slow-roll quantities, $\epsilon$ and $\eta$. Since the inflationary potential itself is very flat and the loop corrections grow as $\delta V_y\propto \phi^4$, they can be subdominant in magnitude but strongly affect $\epsilon$ and $\eta$. We consider the total potential as
\beq
V_{\rm tot} = V(\phi) + \delta V_y(\phi) \, .
\eeq
where $V(\phi)$ is the inflationary potential of Eq.~\eqref{eq:potential}.
The first slow-roll parameter becomes
\beq
\label{eq:epsilontotaly}
\epsilon_{\rm{tot}} \simeq {M^2_{\rm {Pl}}\over 2} \left ({V_{\rm {tot}}'\over V}\right)^2 = 
\left (
\sqrt{\epsilon } 
-
{{\cal N}_f y^4 |\phi|^3 M_{\rm Pl} \over 4\sqrt{2}\pi^2 V} 
\right )^2
\simeq
\left(
\sqrt{ {3\over 4N^2} {\alpha\over M_{\rm Pl}^2} } -
{{\cal N}_f y^4 |\phi|^3 M_{\rm Pl} \over 4\sqrt{2}\pi^2 V} 
\right )^2\,,
\eeq
where $\epsilon$ is the first slow-roll parameter of $\alpha$ attractors given in Eq.~\eqref{eq:epsilonSR} $N$ is the number of $e$-folds measured backwards from the end if inflation. 
We can write the above equation as
\beq
\epsilon_{\rm tot}\equiv (\sqrt{\epsilon}-\sqrt{\delta\epsilon_y})^2\,,
\eeq
and compute the parameter range for $\delta \epsilon_y/\epsilon<1$. Using Eq.~\eqref{eq:phivsN} and $N=60$, we derive
\beq
{\delta \epsilon_y \over \epsilon} =
y^4 {45 {\cal N}_f\over 2\pi^2} {M_{\rm Pl}^2\over\mu^2} \ln ^3\left ({160 M_{\rm Pl}^2\over \alpha}\right )\,.
\eeq
We immediately see that, contrary to $\delta V_y/V$, which is an increasing function of $\alpha$, the ratio $\delta \epsilon_y / \epsilon$ is a monotonically decreasing function of $\alpha$.
We can perform the same exercise with the second slow roll parameter,
leading to
\beq
\label{eq:etatotaly}
\eta_{\rm{tot}} \simeq {M^2_{\rm {Pl}}}  {V_{\rm {tot}}''\over V} = 
\eta +  {3{\cal N}_f y^4 \phi^2 M_{\rm Pl}^2 \over 4\pi^2 V} 
\simeq -{1\over N}+  {3{\cal N}_f y^4 \phi^2 M_{\rm Pl}^2 \over 4\pi^2 V}\,, 
\eeq
where the second slow-roll parameter for $\alpha$-attractors is given in Eq.~\eqref{eq:etaSR}. For $N=60$, $\eta \simeq-0.017$.
Defining
$
\eta_{\rm tot}\equiv \eta+ \delta\eta_y
$
we extract the relative contribution to the second slow roll parameter
\beq
\left |{\delta \eta_y \over \eta}\right | =  {3 {\cal N}_f N y^4 \phi^2 M_{\rm Pl}^2 \over 4\sqrt{2}\pi^2 V}  \, .
\eeq
Figure~\ref{fig:deltaVoverV} shows the monotonically decreasing behavior of $\delta\epsilon_y/\epsilon$ and $\delta\eta_y/\eta$, along with the lines on the $y-\alpha$ plane defining $\delta\epsilon_y = \epsilon$ and $\delta\eta = \eta$. We see that a general requirement for all corrections to be small is 
\beq
\label{eq:yukawabound}
y<{\cal O}(10^{-4}) \, . 
\eeq
This result agrees with that of Ref.~\cite{Ellis:2025bzi}, which computed fermion loop corrections to Starobinsky inflation, a model exhibiting a similar plateau behavior to $\alpha$-attractors.

\bigskip

\subsubsection{Corrections to inflationary potential due to QCD sector}

We now move to examine the corrections to the inflationary potential due to the QCD sector, which are universal regardless of the reheating channel. While Eq.~\eqref{eq:inequalities1} already encompasses the requirement that the energy density of the QCD sector is subdominant to the inflaton potential $\Lambda^4 <V_{\rm {infl}}$, we must also demand that the slow-roll parameters are not altered by the presence of the QCD sector. Since $\Lambda$ depends on $\phi$ and in fact depends exponentially, its derivative with respect to the field  can be important, even if  $\Lambda^4\ll V$. The total potential can be written, similarly to the loop corrections case, as
\beq
{V_{\rm {tot}}(\phi) = V(\phi) -\Lambda^4(\phi)} \, ,
\eeq
where $V(\phi)$ is the inflationary potential of Eq.~\eqref{eq:potential}. 
The negative sign of $\Lambda^4$ corresponds to the fact that the confined phase of QCD has a lower energy density than the de-confined  phase.\footnote{{ Ref.~\cite{vonHarling:2017yew} computes the energy density of the confined QCD sector as $V_{\rm QCD}\simeq-{7\over 17}\Lambda_{\rm QCD}^4$. We neglect the pre-factor for simplicity. Further, in principle we should add today's value of the QCD scale $\Lambda_0^4 \sim (400 \, {\rm MeV})^4$, but we can neglect is as being negligible compared to the high scales of inflation.}}
The first slow-roll parameter becomes
\beq
\label{eq:epsilontotal}
\epsilon_{\rm{tot}} \simeq {M^2_{\rm {Pl}}\over 2} \left ({V_{\rm {tot}}'\over V}\right)^2 = 
\left (
\sqrt{\epsilon } - {44\over \sqrt{2}} {M_{\rm {Pl}}\over M} {\Lambda^4\over V}
\right )^2
\simeq
\left(
\sqrt{ {3\over 4N^2} {\alpha\over M_{\rm Pl}^2} } -
{44\over \sqrt{2}}  {M_{\rm {Pl}}\over M} {\Lambda^4\over V}
\right )^2\,.
\eeq
 Computed at $60$ $e$-folds before the end of inflation the first term of Eq.~\eqref{eq:epsilontotal} becomes $\sqrt{\epsilon}\simeq  \sqrt{2}\times 10^{-2}\sqrt{\alpha}/M_{\rm {Pl}}$
We see immediately that $\Lambda^4<V$ does not guarantee $\epsilon_{\rm {tot}}\simeq \epsilon$. Thus the $\epsilon$ constraint is more restrictive than the energy density one. 

Moving to the second slow-roll parameter
\beq
\label{eq:etatotal}
\eta_{\rm{tot}} \simeq {M^2_{\rm {Pl}}}  {V_{\rm {tot}}''\over V} = 
\eta - 1936 {M_{\rm {Pl}}^2\over M^2} 
{\Lambda^4\over V}
\simeq -{1\over N}- 1936 {M_{\rm {Pl}}^2\over M^2} 
{\Lambda^4\over V}\,,
\eeq
where for $N=60$, $\eta \simeq-0.017$.
Remembering the scaling $M= {\cal O}(1) \sqrt{\alpha}$ we get that the term ${M_{\rm {Pl}}^2 / M^2}$  will be large for $\alpha <M_{\rm {Pl}}^2$, which is the region in which we will operate. We see again that a strong hierarchy between $\Lambda^4$ and $V$ is required to ensure $\eta_{\rm {tot}}\simeq \eta$.

We can re-write the conditions for the slow-roll parameters to be dominated by the inflationary potential instead of the QCD sector in the same form as the ranking of Eq.~\eqref{eq:inequalities}. Requiring that the first term for $\epsilon_{\rm{tot}}, \eta_{\rm {tot}}$ is larger (in absolute value) than the second, we get
\beqn
V^{1/4}>\tilde\epsilon \Lambda\, &,&~
\tilde\epsilon \equiv \left (44 N\sqrt{2\over 3} {M_{\rm {Pl}}^2\over M\sqrt{\alpha}} \right)^{1/4}\,,
\\
V^{1/4}> \tilde\eta \,\Lambda \, &,&~
\tilde\eta \equiv (1936 N)^{1/4} \sqrt{M_{\rm{Pl}}\over M}\,.
\eeqn
The stronger these inequalities become, the closer we get to the pure $\alpha$-attractor results of Eqs.~\eqref{eq:epsilonSR} and~\eqref{eq:etaSR}.

In summary, requiring successful inflation together with controlled loop corrections and subdominant QCD effects imposes non-trivial constraints on the parameter space, particularly on the Yukawa couplings and the hierarchy between the relevant energy scales.

\section{Results for a realization in $\alpha$-attractor inflation models}
\label{sec:results}

This section contains our main results, focusing on the evolution of the QCD confinement scale as controlled by the inflaton dynamics in $\alpha$-attractor models. We begin with the minimal scenario, in which reheating proceeds solely through the inflaton–gluon coupling. In this case, the framework is characterized by two free parameters, $\alpha$
and 
$M$: the parameter 
$\alpha$
 determines the inflaton evolution, while 
$M$
 sets the time at which the QCD confinement scale falls below the Hubble scale, triggering deconfinement. Within this setup, we find that the observed dark matter abundance can be fully generated from de Sitter fluctuations, provided that deconfinement occurs shortly after the CMB modes exit the horizon.

We then consider extensions of this minimal picture. Allowing additional reheating channels, such as inflaton decay into leptons, can increase the reheat temperature but generically requires large Yukawa couplings, leading to significant loop corrections unless they are suppressed, for example through supersymmetry (SUSY). Finally, we adopt a more general approach in which the reheat temperature is treated as a free parameter, enabling a broader exploration of the model’s parameter space. Throughout this section, we use the full expression for the axion variance given in Eq.~\eqref{eq:variancefull}, allowing the relative importance of de Sitter and thermal contributions to be determined dynamically; we indicate in each case which contribution dominates across parameter space.

\begin{figure}
    \centering
    \includegraphics[width=0.45\linewidth]{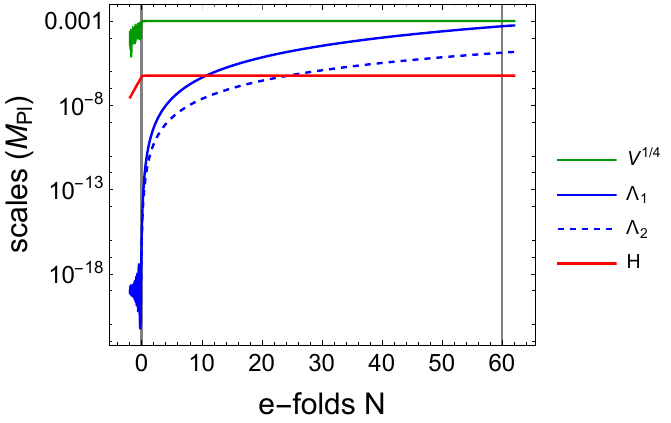}
        \includegraphics[width=0.45\linewidth]{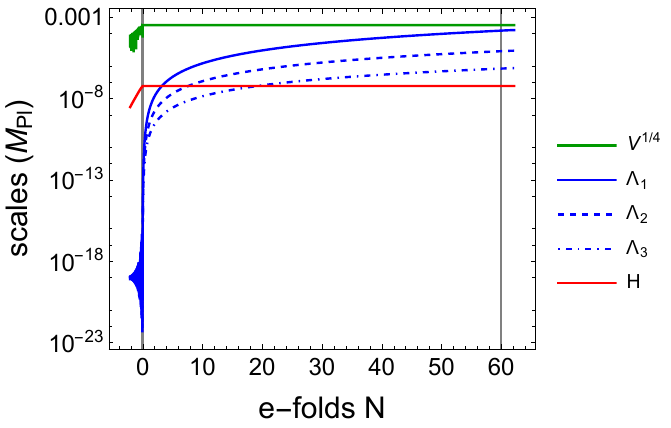}
    \caption{The evolution of the QCD scale in comparison to the Hubble scale (red)  and the potential scale $V^{1/4}$  (green) for $\alpha=0.01M_{\rm Pl}^2$ (left) and $\alpha=0.0001M_{\rm Pl}^2$ (right). The mass-scale $M$ is chosen as follows: for the left panel
     $M/M_{\rm Pl}=0.36,0.4$ (blue solid and dashed  respectively) and for the right panel $M/M_{\rm Pl}=0.055,0.06,0.065$ (blue solid, dashed and dot-dashed  respectively). The vertical gray lines correspond to the end of inflation and the CMB horizon exit moment $N=60$.
     }
    \label{fig:scalesvsN}
\end{figure}
Figure~\ref{fig:scalesvsN} illustrates the evolution of various quantities during inflation: the Hubble scale  $H\sim \mu \sqrt\alpha/\Mp$ (red line), the potential scale $V^{1/4}\sim \sqrt{\mu} \alpha^{1/4}$  (green line), and the QCD scale (blue solid and dashed for $M/M_{\rm Pl}=0.36,0.4$  respectively). The left (right) panels correspond to $\alpha=0.01M_{\rm Pl}^2$  ($\alpha=0.0001M_{\rm Pl}^2$).  
One can see that the two energy scales, $H$ and $V^{1/4}$ are almost constant during inflation, contrary to the QCD scale $\Lambda$, which depends exponentially on the field value $\phi$ and thus flows much more strongly during inflation. It is thus unavoidable (unless we choose $\alpha$ to be so low, as to make this whole discussion moot) that the QCD scale will --at some point during inflation-- become lower than the Hubble scale, signaling a de-confinement. 
At that point, the axion becomes light and acquires de Sitter fluctuations that generate the observed DM abundance in axions (without producing isocurvature modes at CMB scales).
Further the phase transition which will generate a GW signal~\cite{An:2020fff, An:2022cce, An:2023jxf} which is, however, unobservably small, see Section~\ref{sec:GW}). 

Comparing the two panels of Figure~\ref{fig:scalesvsN},
one can see that the ``space" between $H$ and $V^{1/4}$ grows with $\alpha$ as $V^{1/4}/H \propto \alpha^{1/4}$.  The ``distance" between the red and green curves is larger by a factor  $\sim 3$ on the right panel ($\alpha=0.0001\Mp^2$) compared to the left one ($\alpha=0.01\Mp^2$), as $\alpha$ differs by a factor of $100$.

Furthermore,  we wish to point out an intriguing relation  between the mass-scale for the inflaton-gluon coupling and the scale of $\alpha$-attractor inflation.
Following the simple scaling of Eq.~\eqref{eq:phivsN}, we see that $\phi\propto \sqrt{\alpha}$, where we neglected the logarithmic dependence of $\phi$ on $\alpha$. Since the QCD scale is controlled by the exponential of the ratio $\phi/M$, see Eq.~\eqref{eq:LambdaQCD}, we expect that the scale $M$ will share the scaling of $\phi$, meaning that $M\propto\sqrt{\alpha}$. Indeed, if we test the values of Figure~\ref{fig:scalesvsN}, we see that $M/\sqrt{\alpha} = {\cal O}(1)$ for all pairs of parameters. If we include the logarithm of Eq.~\eqref{eq:phivsN}, the match becomes even closer.
The scaling\footnote{If we include the logarithmic dependence of $\phi$, the scaling for $10^{-6}\lesssim \alpha/M_{\rm {Pl}}^2 \lesssim 0.1$  can be better approximated as $M\propto \alpha^{0.4}$.\label{foot:Mscaling}} $M\propto \sqrt{\alpha}$ leads to an intriguing relation  between the mass-scale for the inflaton-gluon coupling and the scale of $\alpha$-attractor inflation, which can inform model-building attempts at unifying inflation with the SM at in a UV-complete theory. While we will not  pursue this relation further in this work, we believe that it is important to note it as an important feature of this model.

\begin{figure}
    \centering
    \includegraphics[width=0.45\linewidth]{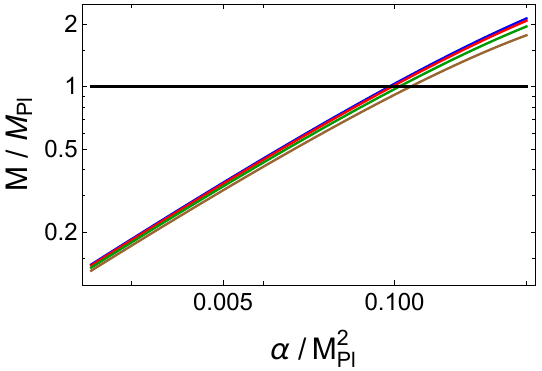}
        \includegraphics[width=0.45\linewidth]{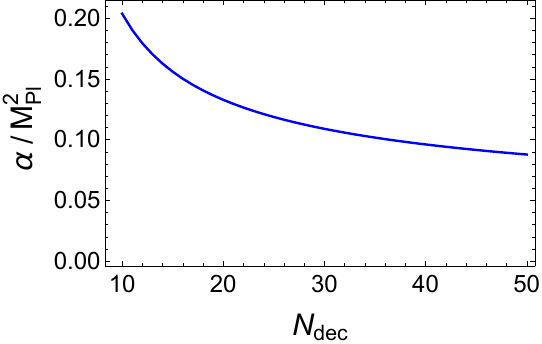}
    \caption{ {\it Left:} The value of the scale $M$ as a function of $\alpha$ for $N_{\rm dec}=45,40,30,20$ (blue,red green and brown respectively). The horizontal black like is a visual cue for the value $M=M_{\rm {Pl}}$.
   {\it Right:} The value of $\alpha$ for which $M=M_{\rm {Pl}}$ as a function of $N_{\rm {dec}}$. Values of parameters below the blue line lead to $M<M_{\rm {Pl}}$. 
    }
    \label{fig:Mvsa}
\end{figure}

As a reminder, we define the deconfinement time $N_{\rm dec}$ as the number of $e$-folds before the end of inflation at which the condition $\Lambda = H$ is satisfied, i.e.\ when the QCD sector transitions from confinement to deconfinement. In this notation, the duration of the light-axion phase is given by $\Delta N \simeq N_{\rm dec}$.
We see in Figure~\ref{fig:scalesvsN} that varying the scale $M$, while keeping all else constant shifts the time when the relation $\Lambda>H$ transitions   to $\Lambda<H$. Figure~\ref{fig:Mvsa} shows the dependence of $M$ on $\alpha$ for different values of $N_{\rm {dec}}$. We see that given $N_{\rm {dec}}$ increasing $\alpha$ beyond some value leads to $M>M_{\rm {Pl}}$. As we wish to keep all scales in the problem below the Planck scale, we define a maximum value of $\alpha$, for which  $M=M_{\rm {Pl}}$. This is only a function of $N_{\rm {dec}}$, as shown by the right panel of Figure~\ref{fig:Mvsa}.

The necessary conditions for the success of this model are listed and explained in Section~\ref{sec:model}. We  list the proper sequence of scales here as a reminder, with the added constraints that the slow-roll parameters $\epsilon, \eta$ are dominated by the inflaton potential 
\beq
\label{eq:inequalities}
\boxed{3H/2<m_a <\Lambda<
\{f_a ,\, V^{1/4}
\} \, , \quad  \{\tilde \epsilon \Lambda,\tilde \eta\Lambda\}<V^{1/4}\, , \quad M<\Mp}  \, .
\eeq

\subsection{Reheating into gluons}
\label{sec:gluons}

The model in its simplest form --where the inflaton couples solely to gluons through a $\phi GG$ term-- has essentially two free parameters: $\alpha$ and $M$ (the Yukawa coupling $y$ will be added in Section~\ref{sec:yukawa}). In principle, another inflationary plateau model can be chosen, which can have more parameters and a different detailed evolution. However $\alpha$-attractors are sufficient to demonstrate the inner workings and limitations of this model and thus we defer a more extensive  exploration of different inflationary realizations for future work. Another note we must make is that the coupling $\phi GG$ corresponds to the lowest order operator and thus generalizations of the form $\phi^n GG$ will lead to quantitative differences, possibly alleviating (or worsening) the limitations of this specific realization. This is also left for future work. 

For the rest of this work, we  keep $\alpha$ as a free parameter and exchange the mass-scale $M$ for $N_{\rm dec}$, the number of $e$-folds before the end of inflation   when     the de-confinement transition occurs. 
As we discussed earlier and is shown in Figure~\ref{fig:scalesvsN}, for a given value of $\alpha$, changing $M$ uniquely determines $N_{\rm {dec}}$.
For a given $N_{\rm {dec}}$, we can then plot all relevant scales as a function of $\alpha$. We start our analysis by using the upper limit for the thermally induced axion variance, $\sigma_{\theta,{\rm thermal}}^2 = T_{\rm reh}^2/f_a^2$.

\begin{figure}[h!]
    \centering
 \includegraphics[width=0.45\linewidth]{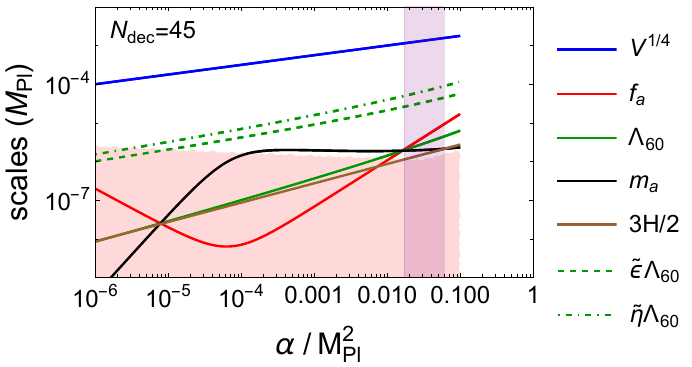}
        \includegraphics[width=0.45\linewidth]{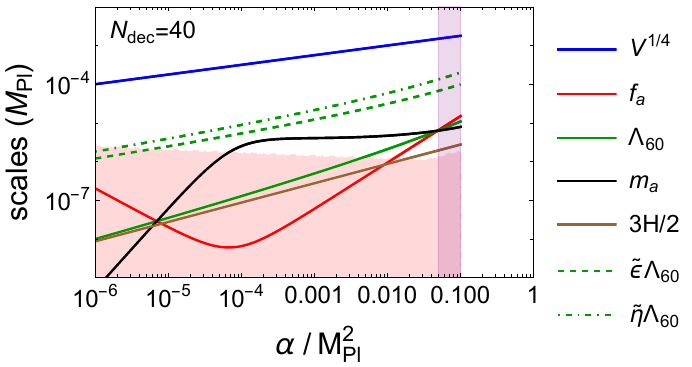}
        \\
          \includegraphics[width=0.45\linewidth]{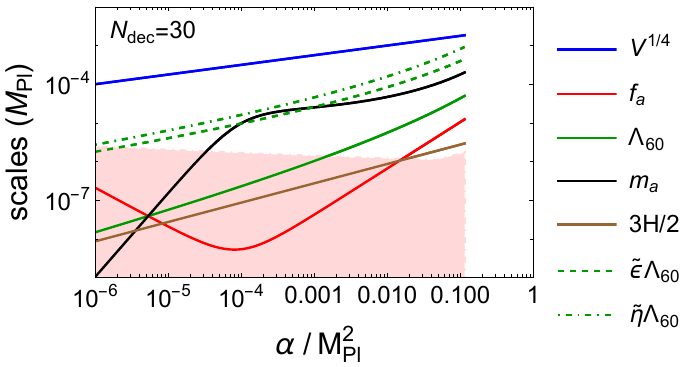}
        \includegraphics[width=0.45\linewidth]{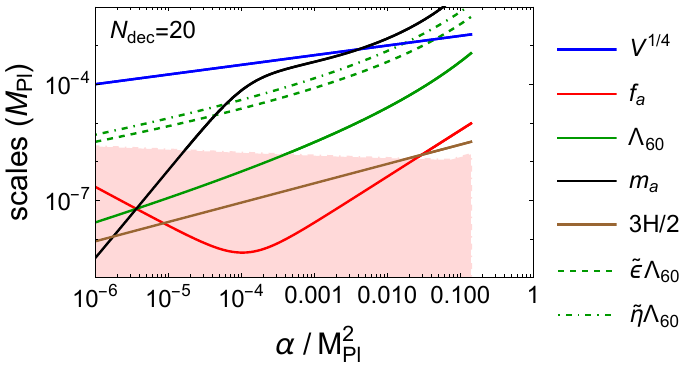}
    \caption{
    The case of minimal reheating, where the inflaton decays to gluons only.
    The dependence of relevant mass/energy-scales is shown as a function of $\alpha$ for $N_{\rm {dec}}=45,40,30,20$ (clockwise from the top left, where $N_{\rm dec}$ is the number of $e$-folds before the end of inflation at which the QCD confinement scale drops below the Hubble scale, triggering deconfinement). In obtaining this figure, we have required that the QCD axion provides 100\% of the DM abundance, as per Eqns.~\eqref{eq:DMabundace} and~\eqref{eq:faDM}; for the thermal contribution, we have
    used the largest allowed  thermally induced axion variance $\sigma_{\theta,{\rm thermal}}^2 = T_{\rm reh}^2/f_a^2$. The top of the pink band shows the value of $f_a$ below which the PQ symmetry is restored during reheating (i.e. $T_{\rm max} > f_a$) and thus if the red line enters the pink band  our DM calculation is invalid. 
  The purple vertical band shows the regime where all inequalities of Eq.~\eqref{eq:inequalities} are satisfied. 
   We consider de Sitter fluctuations and reheating solely through the inflaton-gluon coupling. We see that in this case the model is successful only if QCD deconfines shortly after  the CMB window, which occurs $\sim 60-50$ $e$-folds before the end of inflation. 
   We see that success of the model requires $N_{\rm dec}\gtrsim 40$, i.e. de-confinement must occur soon after the CMB relevant window.
    }
    \label{fig:alphascan}
\end{figure}

The dependence of relevant mass/energy-scales as a function of $\alpha$ is shown in Figure~\ref{fig:alphascan} for four different values of $N_{\rm {dec}}=45,40,30,20$.
For visual clarity we chose the color-coding as follows: In order for the model to be successful, the solid lines must be in the order
in which they appear in the figure legend; from top to bottom: blue, red, green, black and brown. In obtaining this figure, we have required that the QCD axion provides 100\% of the DM abundance, as per Eqs.~\eqref{eq:DMabundace} and~\eqref{eq:faDM}; for the thermal contribution, we have
    used the largest allowed  thermally induced axion variance $\sigma_{\theta,{\rm thermal}}^2 = T_{\rm reh}^2/f_a^2$.
\\
When the axion production is dominated by thermal fluctuations form reheating, which occurs for $\alpha\lesssim 10^{-4}M_{\rm Pl}^2$, the red line (corresponding to $f_a$) is inside the pink band (corresponding to $f_a<T_{\rm max}$). This means that PQ symmetry is restored during reheating and broken again at a later stage. While this does not invalidate our model as a suppression mechanism for axion isocurvature, it does invalidate the DM calculation given by Eq.~\eqref{eq:DMabundace}. This is because after the PQ symmetry breaking we expect the formation of domain walls and cosmic strings, which will eventually decay through emission of axions. Therefore, the final axion abundance cannot be computed through misalignment calculations and needs to be extracted from simulations. For example, Ref.~\cite{Benabou:2024msj} suggests an axion mass of $m_a\sim 40-300\, \mu eV$, leading to an axion decay constant of $f_a\sim 10^{10}-10^{11}\, {\rm GeV}$ or $10^{-9} \, M_{\rm {Pl}}$. This is below the axion decay constant of $f_a\sim 10^{-8}-10^{-7}\, M_{\rm {Pl}}$ seen in the left parts of Figure~\ref{fig:alphascan}.

The vertical purple band in Figure~\ref{fig:alphascan} shows the values of $\alpha$, where the model is successful.
For the correct DM abundance to arise from de Sitter fluctuations, while at the same time keeping the axion decay constant large enough to avoid restoring PQ symmetry, we are pushed toward large values of $N_{\rm {dec}}\gtrsim 40$. Physically, this means that the deconfinement occurs shortly (a few $e$-folds) after the CMB-relevant modes have exited the horizon. 
 Furthermore, we are also pushed to somewhat large values of $0.1\gtrsim \alpha/M_{\rm {Pl}}^2\gtrsim 0.01$. Unfortunately, these values lead to a tensor-to-scalar ratio of $r\lesssim 3\times 10^{-4}$, which is below the sensitivity of current and upcoming CMB experiments, which are expected to probe values of $r\sim 10^{-3}$.

To ensure that the PQ symmetry remains broken after
inflation---so that topological defects are not regenerated and the
axion initial conditions remain well-defined---it is sufficient to
require that the maximum temperature attained after inflation does
not exceed the PQ scale. In the absence of a specific UV completion,
we adopt the  criterion
$T_{max} \lesssim f_a$.

One may also ask whether de Sitter fluctuations during inflation could
stochastically drive the PQ field back toward the symmetric point.
While the Gibbons--Hawking temperature
$T_{\rm GH}=H/2\pi$ does not correspond to a thermal bath and therefore
does not generate the usual plasma-induced restoring term,
a parametrically light PQ radial mode $\rho$ would undergo a random walk
with step size $\delta\rho\simeq H/2\pi$ per e-fold.
If the radial mode is effectively light for $N_{\rm dec}$ $e$-folds,
the resulting rms displacement is
\begin{equation}
\delta\rho_{\rm rms}
\simeq \frac{H}{2\pi}\sqrt{N_{\rm dec}}.
\end{equation}
A conservative sufficient condition to prevent stochastic excursions
toward the origin is therefore
\begin{equation}
\frac{H}{2\pi}\sqrt{N_{\rm dec}} \ll f_a.
\end{equation}
More generally, if the radial mode satisfies
$m_\rho \gtrsim \mathcal O(1)\,H$ during inflation,
its de Sitter fluctuations are suppressed and the PQ symmetry remains
safely broken. To be conservative, we require ${H\over 2\pi}\sqrt{N_{\rm dec}}<f_a$ in  addition to $T_{\rm max}<f_a$. To satisfy both requirements, $f_a$ must lie above the pink band in Figure~\ref{fig:alphascan} and others.

Before we end the discussion of the minimal scenario, where the inflaton is coupled to the SM solely through gluons via the $\phi GG$ operator, it is worth discussing the regimes and parameter dependence of the $f_a$, based on whether the axion variance is dominated by thermal or de Sitter fluctuations. 
For $\alpha\gtrsim 10^{-4}M_{\rm Pl}^2$,  the axion variance (and therefore abundance) is dominated by de Sitter fluctuations. Then $f_a \propto H^{12/5}$, as seen in Eq.~\eqref{eq:faDM}.
Since $H\propto \sqrt{\alpha}$, on the log-log plot of Figure~\ref{fig:alphascan}, $f_a$ is  monotonically increasing  when dominated by de Sitter fluctuations.
For $\alpha\lesssim 10^{-4}M_{\rm Pl}^2$,  the axion variance (and therefore abundance) is dominated by de Sitter fluctuations. The reheat temperature in the case of solely inflaton-gluon coupling is given in Eq.~\eqref{eq:Trehgg} as $T_{\rm{reh},gg} \simeq 0.1 m_\phi\sqrt{m_\phi M_{\rm {Pl}}}/M_*$. Due to the approximate scaling $M_* \propto \alpha^{0.4}$ (see  Footnote~\ref{foot:Mscaling}) the reheat temperature and $f_a$ when dominated by thermal fluctuations  will be a  monotonically decreasing line on a log-log plot.
The thermal and de Sitter contributions to the axion abundance are comparable $\sigma_{\theta,dS} \simeq \sigma_{\theta, {\rm thermal}}$ at the value of $\alpha\simeq 10^{-4}M_{\rm Pl}^2$ where $f_a$ is at its minimum.
The vertical purple band in Figure~\ref{fig:alphascan}, the region where the model is successful, corresponds to larger values of $\alpha$.  Thus, in the minimal reheating scenario described here,  de Sitter fluctuations dominate the axion production.

In summary, the minimal scenario with reheating into gluons successfully suppresses isocurvature perturbations and generates the observed dark matter abundance, 
as long as deconfinement occurs shortly after the CMB relevant window during inflation.  Further, axion production is dominated by de Sitter fluctuations (rather than thermal fluctuations).

\subsection{Additional reheating into fermions}
\label{sec:yukawa}

As yet we have considered only the minimal scenario in which the inflaton-gluon coupling, that raises the QCD scale and axion mass during inflation, is also the only channel for reheating 
after inflation.  In the current section, we consider extensions in which the inflaton couples also to right-handed neutrinos,
allowing for additional reheating channels and potentially higher reheat temperatures. 
First we will show that,  
in the absence of additional structure, the Yukawa couplings required to achieve efficient reheating induce large radiative corrections to the inflaton potential, which can spoil slow-roll dynamics. As a result, this scenario is not viable.
Then we will turn to a successful variant of this model in the context of a supersymmetric setup.

In addition to the two parameters $\alpha$ and $M$ (or $\alpha$ and $N_{\rm {dec}}$),
we must add two more parameters: the Yukawa coupling strength $y$ and the number of right handed neutrinos  coupled to the inflaton ${\cal N}_f$. 
We again start by using the largest possible (most optimistic) value $\sigma_{\theta,{\rm thermal}}^2=T_{\rm reh}^2/f_a^2$ for the thermal excitations of the axion after inflation. In order to ensure that fermion loop corrections do not dominate the potential and slow roll parameters, we need to satisfy the bound on the Yukawa coupling $y<{\cal O}(10^{-4})$ in Eq.~\eqref{eq:yukawabound}.  We chose three representative values of the Yukawa coupling, $y=10^{-5}, 10^{-4},$ and $5\times 10^{-4}$. 
Regarding the remaining free parameters, we choose $N_{\rm dec}=20,30,40$ and ${\cal N}_f=1,3,6$. 
Interestingly, 
for all $27$ parameter combinations, the only ones that gave a viable parameter space are the ones corresponding to $N_{\rm dec}=40$. However, in this case, the allowed parameter space is independent of $y$ and ${\cal N}_f$ and identical to the $N_{\rm dec}$ panel of Figure~\ref{fig:alphascan}. Again, as in the previous section, axion production is dominated by de Sitter rather than thermal fluctuations.

If we  now move away from the overly optimistic  upper bound $\sigma_{\theta,{\rm thermal}}^2=T_{\rm reh}^2/f_a^2$ and use instead the more realistic Hubble-averaged value $\sigma_{\theta,{\rm thermal}}^2 =  H_{\rm reh}T_{\rm reh}/f_a^2$, we again find --as expected-- that the allowed parameter space drives us back to the minimal scenario, where axion production is dominated by de Sitter fluctuations, making the reheating dynamics irrelevant. 
One avenue for circumventing the no-go result for fermion reheating could be to include the fermion loop corrections into the potential and derive the slow-roll evolution of the model. However, fermion loop corrections have a polynomial dependence on the inflaton field, $\Delta V_y\propto \phi^4$. This causes a significant deviation from the flat plateau potential required for our mechanism to work. As we showed in Ref~\cite{Freese:2026xax}, monomial potentials are ruled out, a result which can be extended to potentials with a significant polynomial component. Overall, if the Yukawa couplings are kept small enough, so as not to spoil the inflationary plateau potential, their reheating effects do not affect the axion production, which is dominated by de Sitter fluctuations and thus is the same as in the case of reheating to gluons, described in Section~\ref{sec:gluons}.

Supersymmetry (SUSY) can offer a natural way to control these radiative effects.
Here we use SUSY only as an effective way to control the inflaton-sector radiative corrections induced by the Yukawa couplings; a complete supersymmetric realization of the axion/PQ sector is beyond the scope of this work.
In a supersymmetric theory, every fermion is accompanied by scalar superpartners whose loop contributions enter the effective potential with opposite sign. In the limit of unbroken supersymmetry, the bosonic and fermionic contributions cancel exactly in the one-loop effective potential, eliminating the large Coleman–Weinberg corrections that arise in the non-supersymmetric case. Although inflation necessarily involves some degree of supersymmetry breaking, the resulting mass splittings within supermultiplets can be small enough in suitable embeddings, so that the remaining loop corrections are parametrically suppressed. Consequently, embedding the inflaton sector in a supersymmetric framework can stabilize the inflaton potential against the large radiative corrections generated by Yukawa couplings to right-handed neutrinos. This can open up a parameter space that includes larger Yukawa couplings $y>10^{-4}$.  
In a supersymmetric completion, additional bosonic decay channels can open alongside the fermionic ones, so the total inflaton decay width may receive a model-dependent ${\cal O}(1)$ enhancement relative to the purely fermionic estimate.\footnote{The existence of bosonic degrees of freedom may give rise to parametric resonance and thus efficient preheating. We do not examine this here, and thus treat our results as a conservative lower value of the reheat temperature, reached only through perturbative decay.} Since $T_{\rm reh}\propto \Gamma_\phi^{1/2}$, this translates into a model-dependent ${\cal O}(1)$ increase in $T_{\rm reh}$, and hence in $T_{\max}$ and $\sigma_{\theta,\mathrm{thermal}}$; for definiteness, in the numerical examples below we assume doubling of the degrees of freedom, which increases $T_{\rm reh}$ by factor of $\sqrt{2}$.

Integrating our model in a SUSY framework is beyond the scope of our work. However, we can use an estimate of the loop corrections, due to SUSY breaking at a scale $\tilde m$ (see e.g.~Ref.~\cite{Ellis:2025bzi})
\beq
\label{eq:deltaVSUSY}
\Delta V_{\rm SUSY}  \sim {1\over 16\pi^2} y^2\tilde m^2 \phi^2 
\, .
\eeq
Eq.~\eqref{eq:deltaVSUSY} can be understood as the leading term in the SUSY Coleman--Weinberg supertrace for a split supermultiplet with scalar and fermion masses \(m_s^2 = y^2\phi^2 + \tilde m^2\) and \(m_f^2 = y^2\phi^2\), so that the \(y^4\phi^4\) contributions cancel between bosons and fermions and the remaining correction scales as \(\Delta V_{\rm SUSY} \sim (16\pi^2)^{-1} y^2 \tilde m^2 \phi^2\) up to logarithms, valid in the regime \(y^2\phi^2 \gg \tilde m^2\).
In order to provide some numerical estimates for the loop corrections in case of broken SUSY, we refer to  inflation-induced SUSY breaking~\cite{Dvali:1995fb, Kallosh:2004yh}, where $\tilde m = c_H H_I$. The dimensionless parameter $c_H$ defines the relation between the SUSY breaking scale and the Hubble scale. For simplicity one can take $c_H\sim 1$, but we keep it as a free parameter, in order to examine what values work within our model.  The SUSY-generated potential  becomes \beq
\Delta V_{\rm SUSY}\sim {1\over 16\pi^2} y^2 c_H^2  H_I^2\phi^2 \sim {\cal O}(0.01) y^2 c_H^2 H_I^2\phi^2\,, 
\eeq 
and the relative corrections to the potential and slow-roll parameters  are 
\begin{eqnarray}
    {\Delta V_{\rm SUSY}\over V}&\sim& {\cal O}(0.01) y^2c_H^2 
    {H_I^2\phi^2\over H_I^2 M_{\rm Pl}^2}\sim {\cal O}(0.01) y^2c_H^2  {\phi^2\over  M_{\rm Pl}^2} < c_H^2 \,,
    \\
{\Delta V'_{\rm SUSY}M_{\rm Pl}\over \sqrt{2 \epsilon} \, V}&\sim&  {\cal O}(0.01)y^2c_H^2 
     {H_I^2\phi M_{\rm Pl}\over H_I^2 M_{\rm Pl}^2  \sqrt{\epsilon} }\sim {\cal O}(0.01) y^2c_H^2 {\phi\over  M_{\rm Pl}} {1\over  \sqrt{ \epsilon}} \lesssim c_H^2 \,,
\\
{\Delta V''_{\rm SUSY}M_{\rm Pl}^2\over \eta \, V}&\sim&  {\cal O}(0.01)y^2c_H^2 
     {H_I^2 M_{\rm Pl}^2\over H_I^2 M_{\rm Pl}^2 \eta }\sim {\cal O}(0.01)  {y^2 c_H^2  \over \eta} < y^2< c_H^2 \,.
\end{eqnarray}
The value of $\phi/M_{\rm Pl}$ reduces for smaller $\alpha$, thus the maximum possible value in our case occurs for $\alpha\simeq 0.1M_{\rm Pl}^2$, leading to $\phi/M_{\rm Pl}\lesssim 3$. The correction to the potential thus becomes $ {\Delta V_{\rm SUSY}/  V}\sim {\cal O}(0.1)y^2c_H^2$, which is safely smaller than unity for $c_H\le 1$, even if the Yukawa coupling approaches unity.  The corrections to the slow roll parameters require more work. Since $\eta \sim \sqrt{\epsilon}$, let us examine  the correction to the first slow-roll parameter $\epsilon$, since the correction to $\eta$ follows the same logic and will give the same order-of-magnitude result.
\beq
{\Delta V'_{\rm SUSY}M_{\rm Pl}\over \sqrt{2\epsilon} \, V} \sim {\cal O}(0.01)\times y^2 c_H^2\times
60  {M_{\rm Pl}\over \sqrt\alpha} \ln \left (
{160 M_{\rm Pl}^2\over \alpha }
\right)\,.
\eeq
The above ratio can be larger than unity, depending on both the exact value of the ${\cal O}(0.01)$ pre-factor  and the value of $\alpha$. By taking ${\cal O}(0.01)\times 60\sim 1$, we  derive
a simple constraint for the relative SUSY correction to the first slow-roll parameter to be less than unity
\beq
y\lesssim {0.5\over c_H} {\alpha^{1/4}  \over \sqrt{M_{\rm Pl}}} \, .
\eeq
For large vales of $\alpha/M_{\rm Pl}^2\sim 0.1$, a Yukawa coupling $y\simeq 0.3$ saturates the above bound for $c_H=1$.
For the Yukawa coupling to approach unity with the loop corrections being small, 
 the SUSY breaking scale must be smaller than the inflationary Hubble scale, leading to  $c_H = {\cal O}(0.1)$.\footnote{A similar SUSY breaking scale $\tilde m\lesssim 0.1 H_I$ was given e.g. in Ref.~\cite{Nakayama:2011ri}.} 
While the embedding of our model in a specific SUSY scenario is beyond the scope of this work, it presents an exciting avenue for further study.  
A complete SUSY embedding would also require specifying the PQ sector itself. In particular, the condition $T_{\max} \lesssim f_a$ should then be viewed only as a necessary criterion for PQ non-restoration, since the actual evolution depends on the full PQ scalar sector and on whether thermal- and Hubble-induced masses or $A$-terms drive the PQ fields away from, or back toward, the symmetry-restoring origin. In addition, once SUSY is present, the high-$T_{\rm reh}$ regime can be subject to the usual cosmological constraints from gravitinos and moduli, and, for a supersymmetrized PQ sector, from axino and saxion production and decay. Finally, a complete supersymmetric embedding must also ensure that the additional Yukawa sector does not spoil the inflationary plateau already at tree level through K\"ahler-potential or more general supergravity effects.

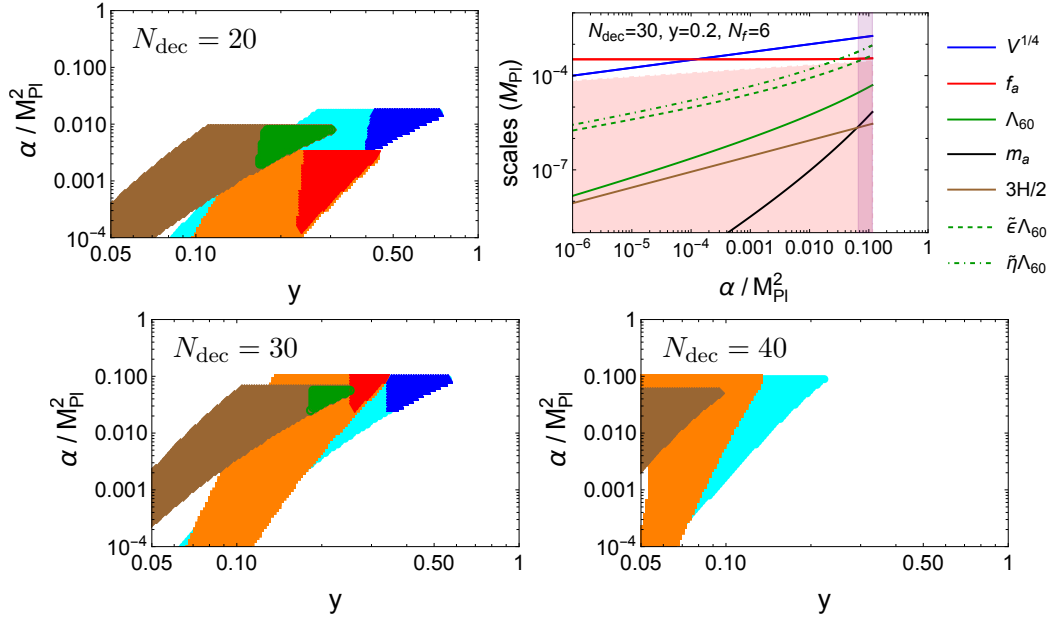
\begin{figure}
      \centering
    
    \begin{tikzpicture}
      \node[anchor=south west, inner sep=0] (img) at (0,0)
        {\includegraphics[width=0.41\textwidth]{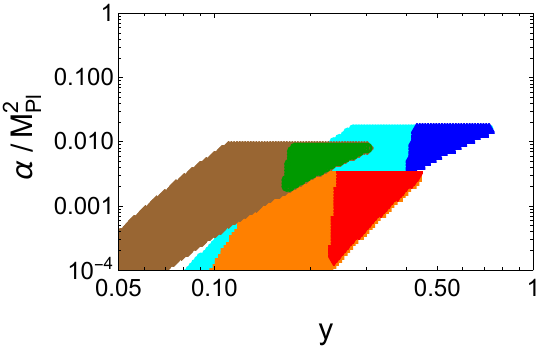}};
      \node at (2.5,3.5) {$N_{\rm dec}=20$};
    \end{tikzpicture}
    \includegraphics[width=0.48\linewidth]{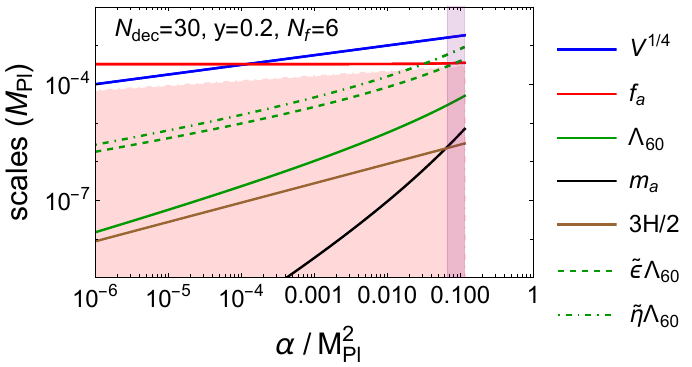}
    \\

    \begin{tikzpicture}
      \node[anchor=south west, inner sep=0] (img) at (0,0)
        {\includegraphics[width=0.41\linewidth]{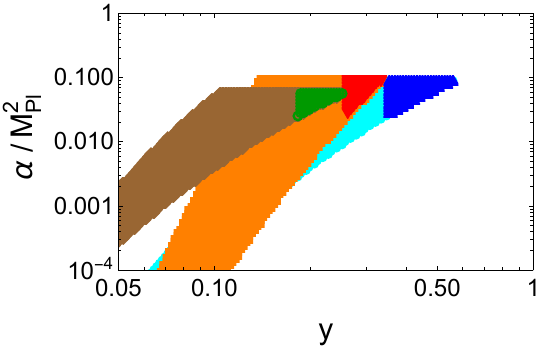}};
      \node at (2.5,3.5) {$N_{\rm dec}=30$};
    \end{tikzpicture}
    %
    \begin{tikzpicture}
      \node[anchor=south west, inner sep=0] (img) at (0,0)
        {\includegraphics[width=0.41\linewidth]{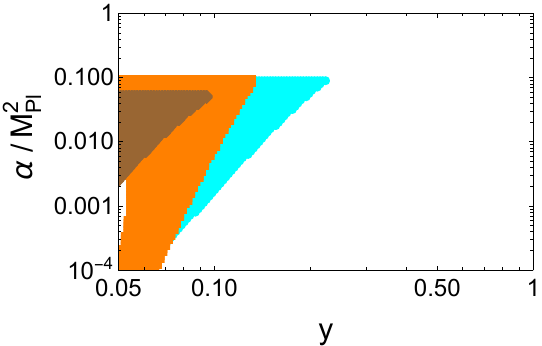}};
      \node at (2.5,3.5) {$N_{\rm dec}=40$};
    \end{tikzpicture}
    \caption{The case of additional reheating into fermions.
The allowed parameter range for $\alpha$ as a function of the Yukawa coupling $y$ is shown 
(satisfying all inequalities of Eq.~\eqref{eq:inequalities}) counterclockwise from the top left panel
for the cases of $N_{\rm dec}=20, 30, 40$ respectively.   
Here the axion production is dominated by thermal fluctuations, where we have used  $\sigma^2_{\theta,{\rm thermal}} = H_{\rm reh}T_{\rm {reh}}/f_a^2$.
 The different colors correspond to different  values of the numbers of right-handed neutrinos ${\cal N}_f=1,3,6$ (cyan/blue, orange/red and brown/green respectively).  The blue, red and green regions  satisfy the PQ non-restoration requirement $f_a>T_{max}$, while the cyan, orange and brown regions violate it.
In order to avoid loop effects, which would invalidate this entire region, we have invoked SUSY to cancel them. For concreteness, we assume that the SUSY completion opens one additional bosonic decay channel per fermionic one, corresponding to a benchmark $\sqrt{2}$ enhancement of $T_{\rm reh}$; the precise enhancement is model dependent.
  {\it Upper right panel:} The dependence of relevant mass/energy scales as a function of $\alpha$  for the case of $N_{\rm dec}=30$, ${\cal N}_f=6$ and $y=0.2$. The color coding for the different bands is the same as in Figure~\ref{fig:alphascan}.   The purple vertical band shows the regime where all inequalities of Eq.~\eqref{eq:inequalities} are satisfied. 
This panel shows an example, where the system satisfies all relevant constraints, if fermion loop corrections are suppressed, as well as provides interesting CMB observables (see Section~\ref{sec:AcT}). 
    }
    \label{fig:fermionsTH}
\end{figure}

{Figure~\ref{fig:fermionsTH} shows the resulting parameter space of the model, when decaying into fermions with large Yukawa couplings $y\gtrsim 0.05$. We have neglected fermion loop corrections --e.g.~by invoking SUSY suppression-- while at the same time doubling the degrees of freedom to which the inflaton decays, thereby counting the super-partners of heavy BSM neutrinos. The upper right panel shows the dependence of relevant mass/energy scales as a function of $\alpha$  for the case of $N_{\rm dec}=30$, $N_f=6$ and $y=0.2$. The color coding for the different bands is the same as in Figure~\ref{fig:alphascan}.  We see from this panel that the allowed values of $\alpha$ are typically large, close to $\alpha\sim 0.1 M_{\rm Pl}^2$. Let us note that larger values $\alpha\gtrsim  0.1 M_{\rm Pl}^2$ lead to $M>M_{\rm Pl}$.  The purple vertical band shows the regime where all inequalities of Eq.~\eqref{eq:inequalities} are satisfied. 
This panel shows an example, where the system satisfies all relevant constraints, if fermion loop corrections are suppressed, as well as provides interesting CMB observables (see Section~\ref{sec:AcT}). 

In the upper right hand panel, the red line corresponds to $f_a$. The top of the pink band shows the value of $f_a$ below which the PQ symmetry is restored during reheating (i.e. $T_{\rm max} > f_a$ for values of $f_a$ within the pink band and thus our DM calculation would be invalid. In this panel, the red line corresponding to $f_a$ is very close to the top of the pink region.
More generally, for all choices of the parameters $N_{\rm dec}$, $N_f$ and $y$, combining all constraints leaves a very small viable region of parameter space for our mechanism to work.
Due to the high reheat temperature, which is required if we want the axion production to be dominated by thermal fluctuations, the maximum temperature during reheating comes close to $V^{1/4}$, leaving a narrow allowed region of values for $f_a$. The remaining allowed parameter space in the $\alpha-y$ plane is shown in the other three panels of Figure~\ref{fig:fermionsTH} (for values of $N_ {\rm dec}$ as labeled) to be  a small triangular region, which depends on the number of fermions ${\cal N}_f$ as shown. Even more interesting is the fact that these triangular regions shrink for larger values of $N_{\rm dec}$ and by $N_{\rm dec}=40$ have vanished entirely. This shows an interesting distinction. In the minimal case of reheating to gluons, where the axions are produced primarily through de Sitter fluctuations, larger values of $N_{\rm dec}\gtrsim 40$ are preferred. In the case of efficient reheating to fermions, leading to a large reheat temperature and thermal-dominated axion production, we are pushed to small values of $N_{\rm dec}\lesssim 30$. 

In the upper left panel and the two lower panels of Figure~\ref{fig:fermionsTH}, we display the viable parameter space both with and without the $T_{\text{max}} < f_a$ constraint. The extended regions (cyan, orange, and brown) illustrate the parameter space that opens up if this constraint is removed, highlighting the severe restriction imposed by  thermal PQ restoration. Taking this constraint seriously is especially critical in our setup; while supersymmetry is required to protect the inflationary plateau from loop corrections, it generally makes the PQ symmetry more susceptible to early-universe restoration due to additional thermal degrees of freedom and positive Hubble-induced masses. Nevertheless, we emphasize that $T_{\text{max}} < f_a$ is a conservative rule of thumb rather than an absolute boundary. The PQ symmetry could potentially survive even if $T_{\text{max}} > f_a$, for example if the peak temperature epoch is  too short-lived to fully drive the scalar field back to the origin.
}

In summary, reheating through fermionic channels requires {BSM physics, for example the addition of right-handed neutrinos with} large Yukawa couplings. Such large  couplings generically spoil inflation through loop corrections, rendering this scenario non-viable in a minimal setup. However, this tension can be resolved in the presence of supersymmetry, which suppresses radiative corrections and restores viability. In particular, loop corrections can be small even for Yukawa couplings approaching unity, if
the SUSY breaking scale is smaller than the  inflationary Hubble by  a factor of ${\cal O}(1-10)$. 

\subsection{Upper limits on the reheat temperature for general reheating (to arbitrary particles)}
\label{sec:generalreheat}

While we have explored the model predictions for the two simplest, natural and well-motivated means of reheating: reheating to gluons via the $\phi GG$ coupling (unavoidable in our model) and reheating to leptons via a Yukawa interaction, it is worth expanding our search to more general means of reheating. For example, coupling of the inflaton $\phi$ to an intermediary scalar or gauge boson field through a trilinear interaction can lead to tachyonic preheating~\cite{Dufaux:2006ee} and thus a much faster transfer of energy from the inflaton than the one achieved through perturbative reheating. 
In preheating the post-inflationary state is generally non-thermal, so the relevant criterion for PQ restoration is controlled by the maximum induced effective mass of the PQ field from the produced excitations. If the system thermalizes rapidly one may equivalently characterize this by an effective maximum temperature 
 defined through the energy density.
Remaining agnostic about the exact reheating mechanism, we will again use~\cite{Giudice:2000ex, Chung:1998rq}
$T_{\rm {max}} =( T_{\rm reh}^2 H_I M_{\rm Pl})^{1/4}$ as the quantity that affects PQ symmetry (non)restoration after inflation.
 The absolute upper limit of the reheat temperature, when instantaneous reheating is considered, is given in Eq.~\eqref{eq:Trehmax}. In the case of $\alpha$-attractors, the Hubble scale at the end if inflation is approximately half of the Hubble scale during inflation $H_{\rm {end}}\simeq 0.5H_I$. We thus get that the instantaneous reheat temperature is 
 \beq
 T_{\rm reh,inst}\simeq 0.4 \sqrt{H_I M_{\rm {Pl}}} \, .
 \eeq
In the case of instantaneous reheating and assuming that the axion variance is dominated by thermal production and not de Sitter fluctuations, the axion decay constant can be computed through the DM abundance of Eq.~\eqref{eq:faDM} as 
\beq
f_a \simeq 
\begin{cases}
50 \left ({\alpha\over M_{\rm {Pl}}^2}\right )^{3/5} & \text{for } \sigma_\theta^2= T_{\rm reh}^2/f_a^2  \\
0.05 \left ({\alpha\over M_{\rm {Pl}}^2}\right )^{9/10}  & \text{for } \sigma_\theta^2= T_{\rm reh}H_{\rm reh}/f_a^2   \, .
\end{cases}
\eeq
Using this value of the axion decay constant and requiring $f_a<M_{\rm {Pl}}$, we  get a constraint on the field-space curvature parameter $\alpha/M_{\rm {Pl}}^2 < 0.0015$  for $\sigma_\theta^2=T_{\rm reh}^2/f_a^2$ and no such constraint for $\sigma_\theta^2=T_{\rm reh}H_{\rm reh}/f_a^2$.
If we take the stricter requirement\footnote{The condition 
$f_a < V^{1/4}$ is not a fundamental requirement, but rather a model-building assumption about 
the PQ sector. If the PQ field arises from a renormalizeable potential 
$V_S \sim \lambda (|S|^2 - f_a^2/2)^2$, then the PQ sector generically 
contributes vacuum energy of order $\lambda f_a^4$. Requiring this 
contribution to remain subdominant during inflation suggests 
$\lambda f_a^4 \ll V$, which for $\lambda = \mathcal{O}(1)$ reduces 
parametrically to $f_a \lesssim V^{1/4}$. Relaxing this assumption 
(e.g.~allowing $\lambda \ll 1$ or other model-dependent suppressions) 
permits $f_a > V^{1/4}$ without spoiling inflation, and does not affect 
the slow-roll or mass-hierarchy constraints discussed in the main text.}
$f_a<V^{1/4}$, 
we arrive at $\alpha/M_{\rm Pl}^2 < 10^{-12}$ for $\sigma_\theta^2=T_{\rm reh}^2/f_a^2$ and $\alpha/M_{\rm Pl}^2 < 0.014$ for $\sigma_\theta^2=H_{\rm reh}T_{\rm reh}/f_a^2$. 
The above constraints refer solely to instantaneous reheating. To fully study the effects of a high reheat temperature on our system, we parametrize the reheat temperature as \beq
T_{\rm {reh}} = c\times T_{\rm reh,inst} \, ~ {\rm with}~c \le 1 \, .
\eeq 
In this parametrization, the dimensionless parameter $c$ encodes the overall reheating efficiency.
With this convention, the axion decay constant becomes, again using Eq.~\eqref{eq:faDM},
\beq
f_a \simeq 
\begin{cases}
50 c^{12/5} \left ({\alpha\over M_{\rm {Pl}}^2}\right )^{3/5} & \text{for } \sigma_\theta^2= T_{\rm reh}^2/f_a^2  \\
0.05c^{18/5} \left ({\alpha\over M_{\rm {Pl}}^2}\right )^{9/10}  & \text{for } \sigma_\theta^2= T_{\rm reh}H_{\rm reh}/f_a^2  
\end{cases}
\eeq
We can now use the same inequalities, $f_a<M_{\rm Pl}$ or $f_a<V^{1/4}$, and constrain $c$ in terms of $\alpha$ 
\beq
\begin{split}
    c&< {0.2 \over (\alpha/M_{\rm {Pl}}^2)^{1/4}} \, , \quad {\rm for}~f_a<M_{\rm {Pl}}~ \&~ \sigma_\theta^2= T_{\rm reh}^2/f_a^2 \,,
\\
c &< {0.018\over {(\alpha/M_{\rm {Pl}}^2)^{7/48}}}\, ,\quad  {\rm for}~f_a<V^{1/4}~ \&~ \sigma_\theta^2= T_{\rm reh}^2/f_a^2  \, ,
\\
c&< {2.3 \over (\alpha/M_{\rm {Pl}}^2)^{1/4}} \, , \quad {\rm for}~f_a<M_{\rm {Pl}}~ \&~ \sigma_\theta^2= H_{\rm reh}T_{\rm reh}/f_a^2 \,,
\\
c &< {0.46\over {(\alpha/M_{\rm {Pl}}^2)^{13/72}}}\, ,\quad  {\rm for}~f_a<V^{1/4}~ \&~ \sigma_\theta^2= H_{\rm reh}T_{\rm reh}/f_a^2  \, .
\end{split}
\eeq
 After inflation, assuming matter domination before reheating, the reheat temperature scales with the scale-factor as $T_{\rm {reh}}\propto a^{-3/8}$. To gain some physical intuition on the parameter $c$, let us take two different examples: a factor of $c\approx 0.1 $  corresponds to about $ 6$ extra $e$-folds of matter-dominated expansion after inflation and a factor of $c\approx 0.5 $  corresponds to about $ 3$ extra $e$-folds.

\begin{figure}[h]
    \centering
   
          \includegraphics[width=0.45\linewidth]{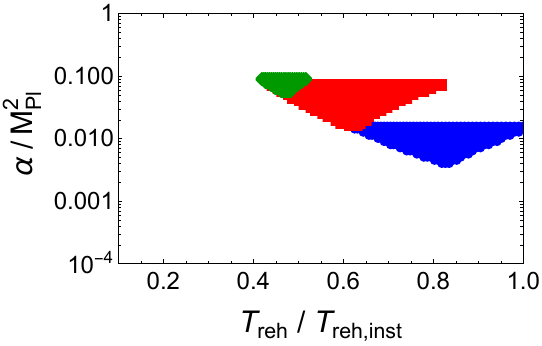}
          \includegraphics[width=0.45\linewidth]{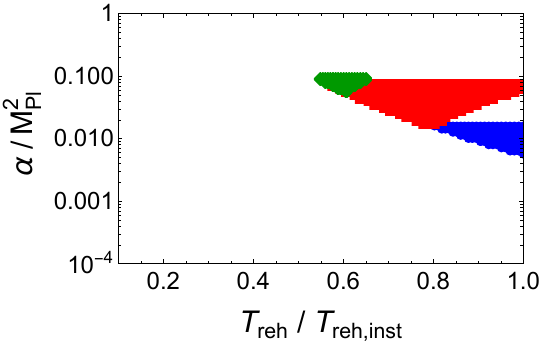}
          \\
          \includegraphics[width=0.45\linewidth]{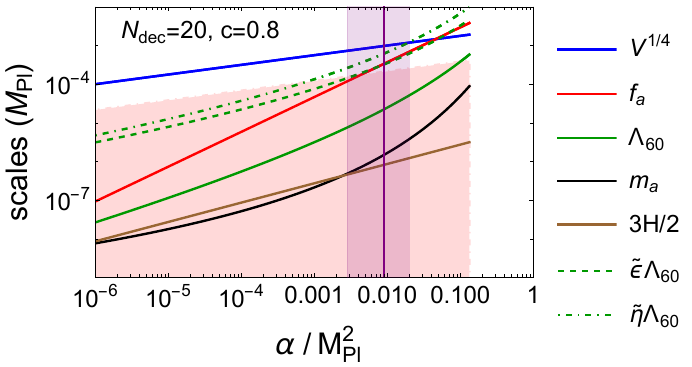}
            \includegraphics[width=0.45\linewidth]{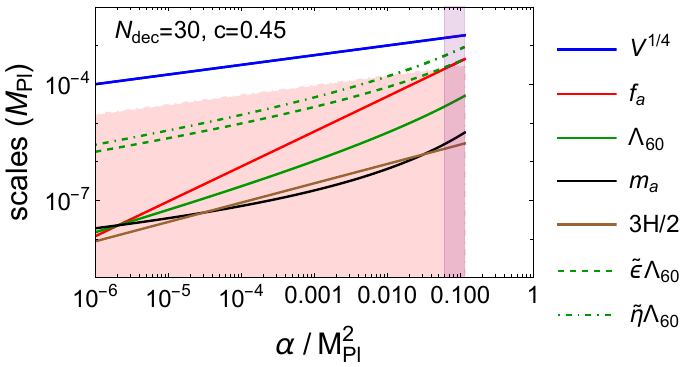}

    \caption{Upper bounds on the reheating temperature for the case of general reheating (to arbitrary particles). For the high values of $T_{\rm reh}$ in this plot, axion production is dominated by thermal fluctuations. {\it Top left:} The allowed region of $\alpha$ as a function of $c\equiv T_{\rm reh}/T_{\rm reh,inst}$, satisfying all inequalities of Eq.~\eqref{eq:inequalities}, as well as $f_a>T_{\rm max}$ for $N_{\rm dec}=20,25,30$ (blue, red and green respectively). The thermally induced axion variance is taken as $\sigma_\theta^2 = H_{\rm reh}T_{\rm reh}/f_a^2$. {\it Top right:} The equivalent plot as the top left panel, but with $\sigma_\theta^2 =0.5 H_{\rm reh}T_{\rm reh}/f_a^2$, thus reducing the interaction strength of the axion with the thermal bath through $\epsilon_{\rm th}=0.5$. 
    {\it Bottom left \& right:} Various mass and energy scales as a function of $\alpha$. Examples are taken from the  case in the upper left panel with $\sigma_\theta^2 = H_{\rm reh}T_{\rm reh}/f_a^2$, specifically choosing $N=20$, $c=0.8$ (left) and $N=30$, $c=0.45$ (right). The vertical purple band shows the viable range of values of $\alpha$.
    The vertical purple line  in the left panel corresponds to a point with significant shift in $n_s$ due to backreaction effects (see Section~\ref{sec:AcT}). 
    }
    \label{fig:alphascan_c}
\end{figure}

{Figure~\ref{fig:alphascan_c} visually summarizes the viability of the model when the reheat temperature, or equivalently $c$, is treated as a free parameter.
The top panels of Figure~\ref{fig:alphascan_c} show the allowed regions in the $\alpha - c$ plane where the model successfully suppresses axion isocurvature perturbations and accounts for 100\% of dark matter, while satisfying all scale inequalities of Eq.~\eqref{eq:inequalities} and the conservative PQ non-restoration bound ($f_a > T_{\text{max}}$).
\\
 The different colors represent different values of $N_{\text{dec}}$. We see that increasing $N_{\text{dec}}$ (i.e., requiring deconfinement to occur earlier) drastically shrinks the allowed parameter space. Eventually, the viable region disappears entirely for $N_{\text{dec}} > 32$.
 By comparing the top-left panel ($\epsilon_{\text{th}} = 1$) to the top-right panel ($\epsilon_{\text{th}} = 0.5$), we see how the efficiency of the axion's thermal interaction affects the model. A weaker interaction with the plasma pushes the allowed parameter space toward larger values of $c$, until the allowed regions of parameter space hit $c=1$ and shrink away.
A note on theoretical consistency: It is worth pointing out a subtle physical tension when choosing $\epsilon_{\text{th}} < 1$ while simultaneously enforcing $f_a > T_{\text{max}}$. If the axion interacts weakly with the plasma ($\epsilon_{\text{th}} < 1$), it is quite possible that the PQ sector never fully thermalizes. In such cases, the strict requirement $f_a > T_{\text{max}}$ for preventing symmetry restoration might be overly conservative. However, we leave it in place to ensure our predictions remain robust and strictly safe from PQ restoration effects. A proper treatment of this subtlety  requires follow-up work,
focusing on specific QCD axion models (e.g. DFSZ or KSVZ) and computing their thermal
history from the end of inflation onward.

The bottom panels of Figure~\ref{fig:alphascan_c} illustrate how the various mass and energy scales evolve as a function of $\alpha$ for two specific benchmark points: $\{N_{\text{dec}}=20, c=0.8\} $ and $\{N_{\rm dec}=30, c=0.45\}$. The vertical purple bands indicate the viable range for $\alpha$ where the correct sequence of scales of Eq.~\eqref{eq:inequalities} is strictly maintained.
A particularly crucial feature appears in the bottom-left panel for the $\{N_{\text{dec}}=20, c=0.8\} $ case. Notice that the upper boundary of the allowed $\alpha$ range (marked by the vertical purple line) is dictated by the intersection of the $\tilde{\eta}\Lambda$ curve (green dashed line) with the $V^{1/4}$ curve (blue solid line). Physically, this boundary represents the point where the QCD sector's backreaction on the inflaton potential becomes large enough to significantly alter the second slow-roll parameter, $\eta$. 

In summary, treating the reheat temperature as a free parameter opens up a broader viable parameter space for the model. We find that successful thermal axion dark matter production, corresponding to high reheat temperature, and isocurvature suppression can be achieved for a variety of reheating histories, provided deconfinement occurs no earlier than roughly 30 $e$-folds before the end of inflation. Furthermore, the upper limits of this newly available parameter space are governed by QCD-induced backreactions on the inflationary slow-roll parameters, setting the stage for interesting CMB phenomenology.
}

\subsection{Modified Spectral Tilt of adiabatic CMB perturbations}
\label{sec:AcT}

In this section, we expand on the fact that the spectral tilt of adiabatic CMB perturbations receives a slight modification towards a smaller value in our scenario.
As we showed in Ref.~\cite{Freese:2026xax}, this effect in inherent given the inflaton-gluon coupling and does not depend on the  specific $\alpha$-attractor realization.

We have shown that for large values of $T_{\rm reh}$, there are combinations of $\alpha$ and $N_{\rm dec}$  where all inequalities
of Eq.~\eqref{eq:inequalities} are satisfied. An intriguing
effect arises from the fact that $\tilde{\eta}\,\Lambda$ approaches $V^{1/4}$
before any other scale (see e.g.~Figure~\ref{fig:alphascan_c}). This means that the second slow-roll parameter $\eta$
can receive a non-negligible contribution from the QCD sector, while all
other scales remain properly aligned according to the inequalities of Eq.~\eqref{eq:inequalities}. 

The spectral tilt of scalar fluctuations, for $\epsilon \ll \eta$ (as is true e.g for $\alpha$-attractor inflation) satisfies
\beq
n_s \simeq 1 + 2\eta \, ,
\eeq
which can be inverted to give
\begin{equation}
\eta \simeq \frac{n_s - 1}{2}.
\end{equation}

The central values for the spectral tilt from the two experiments,
Planck and ACT, are 
 $n_s^{\rm Planck} \simeq 0.965$ and  $n_s^{\rm ACT} \simeq 0.974$,  
leading to the preferred value of the second slow-roll parameter
$\eta^{\rm Planck} \simeq -0.0175$ and $\eta^{\rm ACT} \simeq -0.013$.
Although the difference between the results of Planck and ACT are not of great statistical significance, still it interesting to point out that
the $\alpha$-attractor value (without QCD corrections)
$\eta = -1/N_* \in [-0.0167,-0.02]$ for $N_* \in [50,60]$ is an excellent match to the Planck result but less so to the ACT data. Here $N_*$ is the number of $e$-folds before the end of inflation, when the CMB-relevant modes exited the horizon.
{The correction to $\eta$ due to the QCD potential is negative, as shown in Eq.~\eqref{eq:etatotal}. 
 For
$\alpha\simeq 0.009 M_{\rm {Pl}}^2$, $c=0.02$ and $N_{\rm {dec}}=20$,
the spectral index is shifted by $\delta n_s\simeq -0.01$, taking $n_S$ away from the central value and closer to the left boundary of the ${\it Planck}$ allowed region.}
This parameter choice is shown by the vertical purple line in the lower left panel of Figure~\ref{fig:alphascan_c}. For this choice of parameters, the inflationary Hubble scale is $H_I \simeq 5.5\times 10^{-7}M_{\rm {Pl}}\sim 10^{12}\, {\rm {GeV}} $ and the reheat temperature is $T_{\rm{reh}}\sim  10^{13}\, {\rm {GeV}}$. The same shift in $n_s$ can be also seen in the upper right panel of Figure~\ref{fig:fermionsTH}, where a similar value of $n_s$ is reached for $\alpha\simeq 0.09\, M_{\rm Pl}^2$, when the inflaton decays to 6 fermions and their bosonic superpartners, in a way that suppresses loop effects, with a universal Yukawa coupling $y=0.2$ and the de-confinement occurring $20$ $e$-folds before the end of inflation.

\subsection{Connections to  Cosmology}
\label{sec:overview}

Requiring that all DM is composed of QCD axions, within the context of our model that suppresses isocurvature fluctuations, restricts the early universe to a very narrow range of inflationary Hubble scale $H_I$ and very specific ranges of the reheat temperature $T_{\rm reh}$. The discussion in this section applies to any type of reheating from inflation, whether due to the inflaton/gluon coupling only, the addition of fermionic reheat channels, or any other possible reheat channels, all of which can be characterized by the resulting reheating temperature.

\begin{figure}[h]
    \centering
   
          \includegraphics[width=0.6\linewidth]{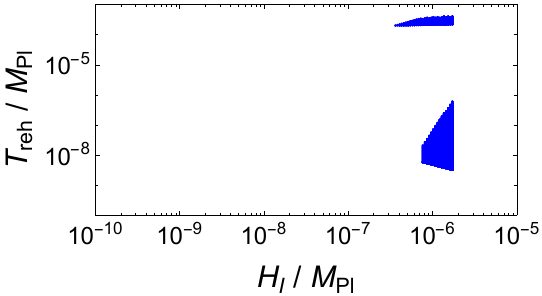}
           
    \caption{
{\it Left:}    The allowed parameter region as a function of the reheat temperature and inflationary Hubble scale, for any reheating type and any de-confinement time $N_{\rm dec}\in [20,45]$, following Section~\ref{sec:generalreheat}. The  lower region corresponds to axion production generated during the last stages of inflation through de Sitter fluctuations, as is relevant for the case of inflaton/gluon coupling only. The upper region corresponds to the axion production generated thermally, as is the case for reheating via additional channels as well and assuming $\epsilon_{\rm th}=1$.
    }
    \label{fig:HTfinal}
\end{figure}

Figure~\ref{fig:HTfinal} shows the allowed parameter range in the $H_I-T_{\rm reh}$ plane for any value of $N_{\rm dec}$, effectively ``integrating over" $N_{\rm dec}$. We see two disjointed regions: one that exhibits a high reheat temperature, and one that appears for lower reheat temperature, below $T_{\rm reh}\sim 10^{-6} M_{\rm Pl}$ and terminates at $T_{\rm reh}\sim 10^{-9}M_{\rm Pl}$. The lower region corresponds to QCD axions  produced  through de Sitter fluctuations and includes reheating solely through gluons production. 
The upper region corresponds to thermally produced axions and requires higher reheating temperatures which necessitate
 reheating through additional channels, such as the right handed neutrinos discussed in this paper.

In all cases we see that successful models require $H_I \sim 10^{-6} M_{\rm Pl}$, corresponding to an inflationary potential height of $V_{\rm inf} \sim 10^{14} - 10^{16}$ GeV or, equivalently, $0.01\lesssim \alpha/M_{\rm Pl}^2\lesssim 0.1$.
These values can also be seen in Figures~\ref{fig:alphascan},~\ref{fig:fermionsTH} and \ref{fig:alphascan_c} above.

Let us examine each of the two ``islands" in Figure~\ref{fig:HTfinal} separately, starting from the bottom of the lower island and moving upwards, along the vertical axis. The bottom part of the lower region corresponds to reheating occurring solely through gluons; a case described in detail in Section~\ref{sec:gluons} and visualized in Figure~\ref{fig:alphascan}.
If we add another reheating channel and start increasing the reheat temperature, initially the axion production will continue to be de Sitter dominated, but  the reheat temperature (and correspondingly the maximum temperature during reheating, $T_{\rm max}$) will increase. This corresponds e.g.~to keeping all the curves in Figure~\ref{fig:alphascan} untouched and only raising the pink region. This is the situation that gives rise to the upper part of the lower region of Figure~\ref{fig:HTfinal}. At some point, the top of the pink region in Figure~\ref{fig:alphascan} ($T_{\rm max}$) will will go above $f_a$, leading to PQ restoration during reheating. In this regime, the reheat temperature is high enough to restore PQ symmetry, but not high enough to generate enough axions thermally to increase $f_a$, per Eq.~\eqref{eq:faDM}. This corresponds to the gap between the two regions of Figure~\ref{fig:HTfinal}.
Further increasing the reheat temperature, the thermal production of axions overcomes the de Sitter production and in turn $f_a$ is raised enough to satisfy $f_a>T_{\rm max}$. This is discussed in Section~\ref{sec:yukawa} and~\ref{sec:generalreheat} and corresponds to the upper region in Figure~\ref{fig:HTfinal}.

As shown previously in Section~\ref{sec:AcT}, the QCD effects during the CMB-observable parts of inflation can lead to shifts in $n_s$.
Specifically, significant shifts in $n_s$
appear only for high reheat temperature in our model, i.e. in the upper region of Figure 9.
In the standard $\alpha$-attractor models in the absence of inflaton/gluon coupling, $n_s = 1 - 2/N_*$.
However, in the presence of the inflaton/gluon coupling of our model, there is additional dependence of $n_s$ not only on $N_*$ but also on the QCD effects, $n_s \equiv n_s(N_*, \Lambda)$.
Indeed, in the context of a plateau model of inflation, with a given value of $N_*$, if the
predicted value of $n_s$ is not in agreement with data, one may speculate that the reason is that QCD effects must be taken into account,
and these require a high reheat temperature.
In this way, precise measurements of the CMB can be thought of as a reheating thermometer.

{Finally, let us note that, while the exact numerical values of Figure~\ref{fig:HTfinal} corresponds to the T-model of $\alpha$-attractors, we expect the results for other plateau models of inflation to be qualitatively similar. }

\section{ Gravitational Waves  from  a (De)confinement transition}
\label{sec:GW}
{In this section, we briefly explore the possibility of generating a stochastic gravitational wave (GW) background from the first-order QCD confinement–deconfinement transition.
 The QCD phase transition has been shown to be first-order for three or more massless dynamical SM quarks~\cite{Pisarski:1983ms}
 (or effectively  massless, i.e. lighter than the QCD confinement scale).
 Therefore, in our scenario, we expect a de-confining first-order phase transition to occur during inflation through bubble nucleation. This process leads to the formation of a stochastic background of GWs exhibiting an oscillatory spectrum~\cite{An:2020fff,Zou:2026wzi}, with the highest peak today at~\cite{An:2020fff}:
\begin{equation}
    f^{\rm peak}_0\sim  10^{11}\,{\rm Hz}\left(\frac{H_I}{M_{\rm Pl}}\right)^{1/2}e^{-(N_{\rm end}-N_{\rm dec})},
\end{equation}
where $N_{\rm end}$ and $N_{\rm dec}$  denote the number of $e$-folds at the end of inflation and when deconfinement takes place (when $H\gtrsim \Lambda(\phi)$), respectively. Unfortunately, the GW spectrum, $\Omega_{\rm GW}$,  scales with the ratio of the energy scale of the QCD sector and the inflationary sector at the time of deconfinement.  Thus $\Omega_{\rm GW}\propto \rho_{\rm QCD}/\rho_{\rm infl}\sim \Lambda^4/V_0$. At the time of deconfinement, $\Lambda\sim H$, leading to $\Omega_{\rm GW}\propto H^4/V_0\sim H^2/M_{\rm {Pl}}^2 \lesssim 10^{-10} $, where we used $V_0\sim H^2M_{\rm {Pl}}^2 $. For a signal coming from a FOPT during inflation to be detectable, Ref~\cite{An:2022cce} uses energy ratios of $\sim {\cal O}(0.1)$, whereas the typical energy ratio in our case will be much less due to the $(H/M_{\rm {Pl}})^2$ suppression, making the signal  undetectable in this model.

\section{Summary and Conclusions}
\label{sec:conclusion}

We have provided a detailed analysis of a simple and predictive framework in which the QCD axion remains consistent with high-scale inflation by dynamically raising the QCD confinement scale during inflation. By introducing a direct coupling between the inflaton and SM gluons~\cite{Freese:2026xax, Dvali:2026ceb}, the strong coupling becomes (exponentially) field-dependent, leading to an enhanced confinement scale $\Lambda(\phi)\gg \Lambda_{\rm QCD}$ during the CMB-relevant epoch of inflation. The resulting axion mass $m_a = \Lambda^2(\phi)/f_a$ is temporarily boosted above the Hubble scale, suppressing de Sitter fluctuations and thereby eliminating the usual inflationary isocurvature bound on $f_a$.

As inflation proceeds, the inflaton rolls toward its minimum and $\Lambda(\phi)$ decreases, inevitably dropping below the Hubble scale before the end of inflation. QCD then undergoes a de-confinement transition, after which the axion becomes light again and acquires fluctuations. These fluctuations can be either de Sitter  fluctuations generated after the CMB-relevant modes have exited the horizon or thermal fluctuations generated during reheating. These fluctuations are the seeds of the axion population that later behaves as cold DM.

We constructed an explicit realization of the mechanism within $\alpha$-attractor inflation models, for which the Hubble scale is nearly constant over the observable $60$ $e$-folds\footnote{The necessity of flat plateau models of inflation was shown in Ref.~\cite{Freese:2026xax}.} and the post-inflation oscillation frequency is insensitive to $\alpha$. The interplay between the exponential sensitivity of $\Lambda(\phi)$ to the inflaton value and the nearly constant inflationary energy scale forces a well-defined hierarchy of physical scales, $3H/2 < m_a < \Lambda < \{f_a , V^{1/4}\}$, in addition to the requirement that the slow-roll parameters are dominated by the inflationary potential, which restrict the viable parameter space. 
If the inflaton-gluon coupling is the only reheat channel, our model requires the de-confinement to occur shortly after the CMB window (around $45$ $e$-folds before the end of inflation), thereby providing a very narrow prediction for the reheat temperature $T_{\rm {reh}}\sim 10^{10}$ GeV and the axion decay constant $f_a\sim 10^{-5} \, M_{\rm {Pl}}\sim 10^{13}\,{\rm {GeV}}$.

In scenarios where reheating proceeds through additional couplings to  right-handed neutrinos, the required large Yukawa couplings generically induce sizable loop corrections to the inflaton potential, which can spoil slow-roll dynamics. As a result, such extensions are not viable in a minimal model,  without some way to suppress loop corrections. This issue can be naturally resolved in the presence of supersymmetry, where cancellations between bosonic and fermionic degrees of freedom suppress the radiative corrections. This allows for larger couplings and higher reheating temperatures, thereby enlarging the viable parameter space of the model while maintaining consistency with inflationary dynamics.  A fully supersymmetric UV completion would also make it possible to revisit the strong-sector running that underlies Eq.~\eqref{eq:LambdaQCD}, since the existence of superpartners to SM particles can modify the $\beta$ function, as well as to compute the detailed PQ thermal history in explicit SUSY DFSZ or KSVZ realizations.

If we remain agnostic to the exact reheating channel and instead use the reheat temperature as a free parameter, the parameter space becomes  broader, as expected. The axion decay constant can reach much higher values, even approaching the Planck scale. 
For high enough reheat temperature, the axion production is dominated by thermal fluctuations and the axion decay constant scales as $f_a\propto T_{\rm reh}^{18/5}$. As $f_a$ increases, the QCD scale and axion mass typically rise as well, and so do the corrections to the slow-roll parameters. 
An example of this can be found in Figure~\ref{fig:alphascan_c} for $T_{\rm {reh}}\sim 10^{13}\,{\rm {GeV}}$, $H_I \sim 10^{12}\,{\rm {GeV}}$ and $N_{\rm {dec}}=20$. There, the model succeeds in suppressing the QCD axion isocurvature perturbations and making the QCD axion the sole dark matter component. Furthermore the corrections to the second slow-roll parameter cannot be neglected.
The extra effect is a non-negligible shift in the spectral tilt $n_s$ coming from QCD back-reaction onto the inflationary potential. This allows us to  use next generation CMB data to constrain the reheat temperature, within the framework of inflaton-driven  QCD confinement.
Finally, the model predicts a first-order confinement/deconfinement transition during inflation, which produces a very high-frequency stochastic gravitational-wave background. Although the amplitude is too small to be observable in foreseeable experiments, the presence of such a signal is a robust outcome of the scenario. 

Overall, this mechanism provides a simple and economical solution to the axion isocurvature problem while preserving the viability of QCD axions as DM with large values of  the axion decay constant ($f_a> 10^{12}$ GeV) and high-scale inflation, with Hubble scale $H \sim 10^{-6} \, M_{\rm {Pl}}$, corresponding to inflationary potential height $V_{\rm inf} \sim 10^{15}$ GeV, as shown in Figures~\ref{fig:alphascan},~\ref{fig:fermionsTH},~\ref{fig:alphascan_c} and~\ref{fig:HTfinal}.
We also found  an interesting distinction. When axions are produced primarily through de Sitter fluctuations (for low enough $T_{\rm reh}$), larger values of $N_{\rm dec}\gtrsim 40$ are preferred, i.e.~deconfinement soon after the CMB window. In the case of efficient reheating, leading to a large reheat temperature and thermal-dominated axion production, we are pushed to small values of $N_{\rm dec}\lesssim 30$.

Our results are robust with respect to several modeling assumptions. In particular, the suppression of axion isocurvature perturbations does not depend sensitively on the precise time dependence of the axion mass, as long as the transition between the heavy and light regimes occurs over a timescale shorter than the duration of inflation. 
Furthermore, one might worry that inflaton quantum fluctuations $\delta \phi$ could introduce a new source of isocurvature by spatially modulating the QCD deconfinement time, $\delta N_{\text{dec}}$. However, because the variation in the e-folding number is exactly the adiabatic curvature perturbation, $\delta N_{\text{dec}} \simeq \zeta$, the resulting spatial variation in the axion abundance is $\delta \rho_a / \rho_a \simeq \zeta / N_{\text{dec}}$. For typical values of $N_{\text{dec}} \sim \mathcal{O}(30)$, this represents a small unimportant  correction to the isocurvature perturbations at small scales, much smaller than CMB observable scales.
Similarly, thermal contributions to the axion abundance are treated parametrically through the efficiency factor $\epsilon_{\mathrm{th}}$, allowing us to capture a wide range of possible interactions between the axion and the plasma without committing to a specific realization. Finally, our criterion for PQ symmetry non-restoration, based on $T_{\max} \lesssim f_a$, is conservative and ensures that our conclusions remain valid even when the thermal history of the PQ sector is model-dependent.

A number of interesting questions arise from the simplicity and success of this model. 
First of all, the exponential dependence of the QCD scale on the inflaton field displacement makes it important to study more inflationary models, in order to examine which of the predictions that appear in this work are particular to the case of the T-model of $\alpha$-attractors and which are generic for this type of $\phi G_{\mu\nu} G^{\mu\nu}$ coupling. Natural extensions include the E-model, polynomial $\alpha$-attractors~\cite{Kallosh:2022feu} and non-minimally coupled models, all of which provide a very stable Hubble scale (which is a prerequisite) with differences in the field evolution.
While we have shown the importance of plateau models in Ref.~\cite{Freese:2026xax}, the detailed evolution of $\phi(N)$ remains model dependent and can alter the available parameter space. 
Different (higher-order) interaction terms of the form $\phi^n/M^n G_{\mu\nu}G^{\mu\nu}$ can also lead to different results. These investigations are necessary to map the model and parameter space of this mechanism for suppressing axion isocurvature perturbations while keeping the QCD axion as a viable DM candidate. Furthermore, the minimal realization of this mechanism within the T-model, where the inflaton gluon coupling is the sole reheating channel, provides an intriguing avenue for a more detailed exploration. In particular, depending on the exact time of the de-confining transition and the exact range of the CM window, the mass of the axion can be comparable or slightly smaller than the Hubble scale during the last part of the CMB window. This opens up the possibility of a very blue isocurvature spectrum, that is essentially only visible in the high CMB multipoles. This can provide a smoking gun signal for this minimal model, in light of dedicated CMB experiments like the Simons Observatory, who will provide particularly 
    high quality date for large multipoles.
   Similar effects can arise if the deconfinement transition occurs at scales relevant to $21$~cm experiments~\cite{Minoda:2021vyw,Sekiguchi:2013lma}.
Furthermore, on even smaller scales, the highly scale-dependent nature of this spectrum can yield $\mathcal{O}(1)$ fluctuations that seed axion miniclusters or ultracompact minihalos. A detailed analysis of these detectable signatures from this model will be presented separately in a follow-up paper.

\acknowledgments
 We are grateful to Gustavo Marques-Tavares and Mark Hertzberg for extremely helpful conversations. K.F. holds the Jeff \& Gail Kodosky Endowed Chair at the University of Texas, Austin. K.F. and E.I.S. are
grateful for support from this Chair. K.F. and E.I.S. acknowledge support from the U.S. Department of
Energy, Office of Science, Office of High Energy Physics program under Award Number DESC-0022021. K.F. also acknowledges support from the Swedish Research Council (Contract
No. 638-2013-8993). B.S.E. is particularly grateful to Pearl Sandick for her continuous support and
encouragement.

\appendix

\section{Axion de Sitter fluctuations}
\label{app:stochastic}

The same physics that was described in Section~\ref{sec:deSitter} can be understood in the stochastic inflation framework. As long as a spectator field (in our case the QCD axion) is heavy during inflation, it is driven exponentially towards its origin. 
 When it becomes light, it will start experiencing de Sitter fluctuations, which displace it from its minimum. We calculate the ``typical" value of the axion displacement following~\cite{Starobinsky:1994bd}.
For simplicity, let us  Taylor-expand the cosine potential and only keep the quadratic (mass) term
$
V \simeq   \frac{1}{2}m_\theta^2 f_a^2\theta^2 $, 
where $m_\theta = \Lambda^2/f_a$. 
The equation of motion in de Sitter space (which is an excellent approximation during plateau inflation) is
\begin{equation}
\ddot \theta - {1\over a^2}\nabla^2 \theta + 3H\dot \theta + m_\phi^2 \theta= 0.
\end{equation}
The gradient terms vanish and  we get, following Ref.~\cite{Starobinsky:1994bd},
\begin{equation}
- f_a^2 {\partial \over\partial  N}\langle  \theta^2\rangle = {H^2\over 4\pi^2} 
- \frac{2 m_\phi^2}{3H^2} f_a^2 \langle  \theta^2\rangle.
\end{equation}
where we measure the number of $e$-folds $N$ from the end of inflation.
Therefore, $ N$ is an decreasing quantity during inflation. The above equation applies only wheh the axion is light.
If the field starts at $\langle \theta^2\rangle 
\simeq 0 $, the evolution is dominated by the $H^2/4\pi^2$ term of the above equation and thus it initially  grows as $(H^2/ 4\pi^2 )\Delta N$ for $m_\phi \ll H$, where $\Delta N$ corresponds to the number of $e$-folds when the axion is light. By solving the above differential equation and neglecting the mass term, the variance as a function of the number of $e$-folds  becomes
\beq
\label{eq:deSittervariance}
\sigma_\theta^2 \equiv {\langle \theta^2\rangle}= {H^2\over 4\pi^2 f_a^2} \Delta N\, .
\eeq
We see that we recover, as expected, the equation that was derived using the approximate axion power-spectrum in Eq.~\eqref{eq:thetavarianceuk}

For $m_a \neq 0$, the field fluctuations would eventually relax to a stationary distribution with variance
\begin{equation}
\langle \theta^2 \rangle_{\rm eq} =
\frac{3 H^4}{8\pi^2 m_a^2 f_a^2}.
\end{equation}
However, for $m_a\ll H$ this would take longer than $60$ $e$-folds and is thus irrelevant to our calculation. We will thus use Eqs.~\eqref{eq:deSittervariance} or equivalently~\eqref{eq:thetavarianceuk} to quantify the axion variance from de Sitter fluctuations during the late stages of inflation. Once the axion fluctuations re-enter the horizon, they can be described as particles with energy density $\rho_a\sim {1\over 2} m_a^2 f_a^2 \sigma_\theta^2$.

\section{Parametric resonance}
\label{app:preheat}

The inflaton-gluon coupling and the bosonic nature of the gluons allows for parametric resonance, which can in principle lead to efficient energy transfer from the inflaton to gluons. We thus need to explore the possibility of complete preheating, the transfer of the entirety of the energy density of the inflaton through parametric resonance, because this would lead to a significantly larger reheat temperature than the one computed using perturbative reheating via Eq.~\eqref{eq:Trehgg}. 
Our goal is not to exactly compute the preheating dynamics, which requires lattice simulations that are challenging for non-Abelian fields, but rather attempt to estimate the magnitude of the energy transfer from the inflaton to gluons through parametric resonance. For that, we  neglect the non-Abelian nature of the gluons and follow closely the derivation of Ref.~\cite{Adshead:2023qiw}, which provides  the equations of motion for a $U(1)$ gauge field coupled to a scalar through a term of the form $-{1\over 4}W(\phi) F_{\mu\nu}F^{\mu\nu}$. This is exactly mirrored in our case, when identifying $W(\phi) = {1/g_{s0}^2 + \phi/M}$. By approximating $G_{\mu\nu}\simeq \partial_\mu A_\nu - \partial_\nu A_\mu$, the linearized equations of motion for the orthogonal polarizations $A_\pm$ and the time-like component $A_0$ (which can be swapped for $A_\parallel$) are written in  Fourier space as
\begin{eqnarray}
    \ddot A_\pm + \left (H + {\dot W\over W}\right ) \dot A_\pm  +
{k^2\over a^2} A_\pm =0,
\\
\ddot A_0 + 5 H \dot A_0 +\left(
{k^2\over a^2} + 3\dot H + 6H^2
\right ) A_0=0.
\end{eqnarray}
These are modified if the gauge field has a mass term, which we do not include for the gluons. We see that the $A_0$ equation of motion does not depend on the coupling to the inflaton field and thus will not experience any parametric resonance. We focus on the equation for $A_\pm$. Using the field redefinition ${\cal A}_\pm = \sqrt{aW} A_\pm$ we arrive at
\beq
\ddot {\cal A}_\pm + \left ( {k^2\over a^2}
-{\partial_t^2 \sqrt{W}\over \sqrt{W}} - {H \dot W\over 2W} - {\dot H\over 2} - {H^2\over 4}
\right ){\cal A}_\pm=0.
\eeq
The longitudinal component is $A_\parallel = ( \sqrt{a}k)^{-1} \left ( \dot{\cal A}_0 + {H\over 2}{\cal A}_0\right )$ and thus will not be excited, following $A_0$.
At the end of inflation, the inflaton will experience damped oscillations. In particular, the potential near the origin is
\beq
V(\phi) \simeq \mu^2  \frac{\phi ^2}{6  }-\mu^2 \frac{\phi ^4}{54 \alpha} + {\cal O}(\phi^6).
\eeq
Since inflation ends at $\phi ={\cal O}\sqrt{\alpha}$ we see that even if the field probes the higher order terms in the potential right at end of inflation, since the field amplitude decays as $\phi\propto a^{-3/2}$, the quadratic approximation will quickly become sufficient.  Of course the quadratic approximation is not able to capture inherently non-linear effects, like the formation of oscillons, however it is sufficient to provide us with an estimate of the parametric excitation of gluon fields.  We can now use the inflaton field value
\beq
\phi(t) = \phi_0 \cos(\mu t/\sqrt{3}) \, ,~ \phi_0\propto a^{-3/2}.
\eeq
In order to construct the Floquet instability diagram, we neglect the Hubble expansion and use the sinusoidal approximation for the infaton motion, leading to
\beq
\ddot {\cal A}_\pm + \left ( {k^2\over a^2}
+\frac{g_{s0}^4 \dot \phi^2-2 g_{s0}^2 \ddot \phi  \left(g_{s0}^2 \phi +M\right)}{4 \left(g_{s0}^2 \phi +M\right)^2}
\right ){\cal A}_\pm=0.
\eeq
Using the fact that $M\simeq 5\sqrt{\alpha}$ and $\phi_{\rm {end}}\simeq \sqrt{\alpha}$, we can simplify the equation of motion of ${\cal A}_\pm$, keeping only the dominant sinusoidal term
\beq
\ddot {\cal A}_\pm + \left ( {k^2} - {g_{s0}^2 \ddot\phi \over 2M}\right ){\cal A}_\pm\simeq0, 
\eeq
which becomes, by using the approximate sinusoidal form of $\phi(t)$
\beq
\ddot {\cal A}_\pm + \left ( {k^2} + {g_{s0}^2 \phi_0 \mu^2\cos(\mu t/\sqrt{3} )\over 6M}  \right ){\cal A}_\pm\simeq0 .
\eeq
We can also re-scale time as $\mu t\sqrt{3}  =\tau$, which leads to
\beq
\partial_\tau^2 {\cal A}_\pm + \left ( {k^2} + {g_{s0}^2 \phi_0  \over 2M}\cos(\tau) \right) {\cal A}_\pm\simeq0. 
\eeq
Since $g_{s0}, \phi_0/M < 1$, we immediately see that there is no significant parametric resonance from the inflaton into gauge bosons. Even if some wave-numbers do start inside the instability band, they will quickly (within one $e$-fold) exit it, meaning no significant enhancement can occur. The Floquet instability chart is shown in Figure~\ref{fig:gluonfloquet}, where we see that only the first instability band is accessible for our regime of $\phi_0/M < 1$. As expected from the kinematics of a two-body decay process ($\phi \to gg$), the tip of this instability band is located precisely at $k = m_\phi/2$, or  $k^2/m_\phi^2 = 0.25$.

\begin{figure}
    \centering
    \includegraphics[width=0.6\linewidth]{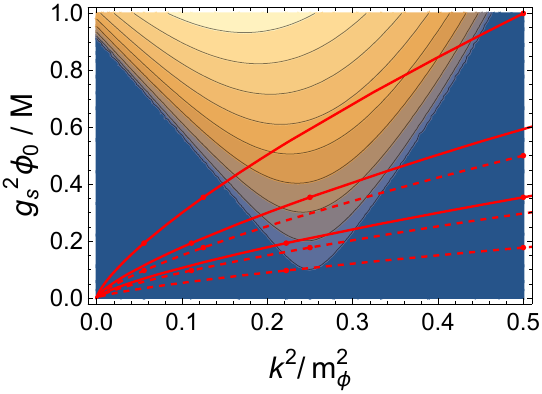}
    \caption{
    The instability chart for parametric excitation of gluons from an oscillating inflaton. The red lines correspond to different $k$-modes as they red-shift and ``flow" through the instability band. The three solid (dashed)  curves correspond to three different vales of $k$. The difference between solid and dashed curves is a different choice of initial $g_s^2 \phi_0/M$, with dashed having a value of $\sim 0.5$ and solid a value of $\sim 1$. The dots on each curve correspond to time intervals of one $e$-fold. The wavenumber is measured in units of $m_\phi$. 
    }
    \label{fig:gluonfloquet}
\end{figure}

In addition to the weak parametric resonance we expect due the fact that the quality parameter $q\equiv g_{s0}^2\phi_0/M <1$, other factors can suppress parametric resonance. Firstly we can estimate the scattering rate of fluons into SM quarks $gg\to q\bar q$ as $\sigma_{gg\to q\bar q}\sim \alpha_s^2/s$ where $s$ is the center of mass energy. The total rate of these effects can be estimated as $\Gamma_{gg\to q\bar q} \sim \sigma v n \sim \sigma \rho_{g}/k \sim \alpha_s^2 \rho_g / m_\phi^3$, where we used $\sqrt{s},k\sim m_\phi$ and the velocity is unity (speed of light). We must compare this to the typical time-scale of parametric resonance, which can be either  $m_\phi$ or $\mu_k\lesssim 0.5 m_\phi$. We see that $\Gamma / m_\phi \sim \alpha_s^2 \rho_g / m_\phi^4$. So when the energy density of the produced gluons reaches  $\rho_{g,{\rm scat}}\gtrsim m_\phi^4/\alpha_s^2\sim 400 m_\phi^4$, gluon scattering into quarks starts depleting the gluon population, thereby  suppressing Bose enhancement. Since the energy density of the inflaton is $\rho_\phi \sim V_0 \sim \alpha m_\phi^2 \gg 100 m_\phi^4$. We estimate $\rho_\phi>\rho_{g,{\rm scat}}$ for $\alpha/M_{\rm {Pl}}^2 \lesssim 10^{-10}$, so without our parameter range, the entirety of the inflaton cannot be transferred to gluons before gluon scattering into quarks starts being efficient. Furthermore, we approximated gluons as Abelian fields, in order to estimate the efficiency of parametric resonance. However, once their occupation number starts increasing,  non-Abelian interactions start producing significant effects, like affective mass terms $m_{\rm {eff}}\sim g^2 \langle A^2\rangle$, which can suppress resonance. This was actually shown in Ref.~\cite{Adshead:2017xll}, where the authors simulated an inflaton with a quadratic potential, coupled  to an $SU(2)$  sector via dilaton-like coupling and found that for a non-Abelian coupling strength larger than ${\cal O}(10^{-3}$ preheating is drastically suppressed.

To summarize, parametric resonance in our model is not efficient, since the ratio of the inflaton amplitude to the mass-scale $M$ at the end of inflation is less than unity. Even if, in a different model, that ratio could be brought up to unity, gluon scattering to quarks and gluon non-Abelian effects are expected to suppress parametric resonance, making the perturbative reheat temperature we computed a universal expectation of the minimal realization of this mechanism (referring to inflaton coupling solely to gluons and no other SM particles).

\bibliographystyle{JHEP}
\bibliography{draft.bib}
\end{document}